\documentclass[sigplan,screen,nonacm]{acmart}
\pdfoutput=1

\AtBeginDocument{%
  \providecommand\BibTeX{{%
    \normalfont B\kern-0.5em{\scshape i\kern-0.25em b}\kern-0.8em\TeX}}}

\setcopyright{rightsretained}
\copyrightyear{2022}
\acmYear{2022}
\acmDOI{}

\acmConference[QPL2022]{Quantum Physics and Logic 2022}{June 27--July 1, 2022}{Oxford, UK}
\acmBooktitle{Proceedings of Quantum Physics and Logic 2022, June 27--July 1, 2022, Oxford, UK}
\acmPrice{}
\acmISBN{}

\usepackage{amsmath}
\usepackage{amsthm}
\usepackage{braket}
\usepackage{enumitem}
\usepackage{float}
\usepackage{hyperref}
\usepackage{mathtools}
\usepackage{microtype}
\usepackage{relsize}
\usepackage{rotating}
\usepackage{tikz-feynman}
\tikzfeynmanset{compat=1.1.0, warn luatex=false}
\usepackage{tikzit}
\usepackage{simpler-wick}
\usepackage[compact]{titlesec}
\usepackage{upgreek}
\usepackage{ulem}
\usepackage{xargs} 
\allowdisplaybreaks

\usetikzlibrary{external}
\newcounter{todocounter}  \newcommandx{\todocount}[2][1=]{\stepcounter{todocounter}\todo[linecolor=YellowGreen,backgroundcolor=YellowGreen!25,bordercolor=YellowGreen,#1]{\thetodocounter: #2}}

\newcommandx{\unsure}[2][1=]{\todo[linecolor=red,backgroundcolor=red!25,bordercolor=red,#1]{#2}}
\newcommandx{\change}[2][1=]{\todo[linecolor=blue,backgroundcolor=blue!25,bordercolor=blue,#1]{#2}}
\newcommandx{\info}[2][1=]{\todo[linecolor=Green,backgroundcolor=Green!25,bordercolor=Green,#1]{#2}}
\newcommandx{\dothis}[2][1=]{\todo[linecolor=Plum,backgroundcolor=Plum!25,bordercolor=Plum,#1]{#2}}

\newcommand{\p}{\partial}
\renewcommand{\vec}[1]{\underline{#1}}
\newcommand{\defeq}{\vcentcolon=}
\newcommand{\lagr}{\mathcal{L}}
\DeclarePairedDelimiter\floor{\lfloor}{\rfloor}
\def\fcmp{\mathbin{\raise 0.6ex\hbox{\oalign{\hfil$\scriptscriptstyle      \mathrm{o}$\hfil\cr\hfil$\scriptscriptstyle\mathrm{9}$\hfil}}}}
\newcommand{\goodchi}{\protect\raisebox{2pt}{$\chi$}} 
\newcommand{\goodtau}{\tau}

\newcommand{\greenSpider}{\!\vcenter{\hbox{\scalebox{0.7}{\begin{tikzpicture}
	\begin{pgfonlayer}{nodelayer}
		\node [style=green dot] (0) at (0, 0) {};
	\end{pgfonlayer}
\end{tikzpicture}
}}}\!}
\newcommand{\redSpider}{\!\vcenter{\hbox{\scalebox{0.7}{\begin{tikzpicture}
	\begin{pgfonlayer}{nodelayer}
		\node [style=red dot] (0) at (0, 0) {};
	\end{pgfonlayer}
\end{tikzpicture}
}}}\!}

\newcommand{\setDefinition}[2]{\left\{#1 \,\middle|\, #2 \right\}}
\newcommand{\naturals}{\mathbb{N}} 
\newcommand{\integers}{\mathbb{Z}} 
\newcommand{\reals}{\mathbb{R}} 
\newcommand{\complexs}{\mathbb{C}} 
\newcommand{\integersMod}[1]{\mathbb{Z}_{#1}} 
\newcommand{\nonstd}[1]{\!\,^\star #1}
\newcommand{\starNaturals}{\nonstd{\naturals}} 
\newcommand{\starComplexs}{\nonstd{\complexs}} 
\newcommand{\starReals}{\nonstd{\reals}} 

\newcommand{\stdpartSym}{\operatorname{st}}
\newcommand{\stdpart}[1]{\stdpartSym\left(#1\right)}

\newcommand{\starIntegersMod}[1]{{\nonstd{\integersMod{#1}}}}
\newcommand{\starIntegersModPow}[2]{{\nonstd{\integersMod{#1}^{#2}}}}

\newcommand{\fHilbCategory}{\operatorname{fHilb}} 
\newcommand{\starHilbCategory}{^\star\!\fHilbCategory} 

\theoremstyle{definition}

\theoremstyle{definition}

\newcommand{\SpaceH}{\mathcal{H}}

\tikzstyle{green dot}=[fill={rgb,255: red,216; green,248; blue,216}, draw=black, shape=circle]
\tikzstyle{red dot}=[fill={rgb,255: red,232; green,165; blue,165}, draw=black, shape=circle]
\tikzstyle{small box}=[fill=white, draw=black, shape=rectangle, minimum width=0.75cm, minimum height=0.75cm, font={\LARGE}]
\tikzstyle{long box}=[fill=white, draw=black, shape=rectangle, minimum width=0.75cm, minimum height=1.25cm, font={\LARGE}]
\tikzstyle{very long box}=[fill=white, draw=black, shape=rectangle, minimum width=0.75cm, minimum height=1.75cm, font={\LARGE}]
\tikzstyle{state}=[draw, rounded rectangle, rounded rectangle east arc=0pt, inner sep=2pt, minimum height=0.6cm, minimum width=0.8cm, fill=white, font={\LARGE}]
\tikzstyle{effect}=[draw, rounded rectangle, rounded rectangle west arc=0pt, inner sep=2pt, minimum height=0.6cm, minimum width=0.8cm, fill=white, font={\LARGE}]
\tikzstyle{split}=[fill=black, draw=none, regular polygon, regular polygon sides=3, rotate=90, minimum width=3mm, inner sep=2pt]
\tikzstyle{merge}=[fill=black, draw=none, regular polygon, regular polygon sides=3, rotate=-90, minimum width=3mm, inner sep=2pt]
\tikzstyle{diamond}=[fill=white, draw=black, regular polygon, regular polygon sides=4, shape border rotate=45, inner sep=1pt, minimum height=0.8cm, minimum width=0.8cm, font={\LARGE}]
\tikzstyle{white square}=[fill=white, draw=black, shape=rectangle, font={\LARGE}]
\tikzstyle{small state}=[draw, rounded rectangle, rounded rectangle east arc=0pt, inner sep=2pt, minimum height=0pt, minimum width=0pt, fill=white]

\tikzstyle{dashed}=[-, dash pattern=on 2mm off 1mm]
\tikzstyle{blue}=[-, draw=blue]

\newcommand{\SpaceX}{\mathcal{X}}
\newcommand{\SpaceT}{\mathcal{T}}
\newcommand{\tikzfigscale}[2]{\ensuremath{\vcenter{\hbox{\scalebox{#1}{$\input{#2.tikz}$}}}}}

\begin{document}

\title{Categorical Semantics for Feynman Diagrams}


\author{Razin A. Shaikh}
\authornote{Both authors contributed equally to this research.}
\orcid{0000-0001-8995-5898}
\affiliation{%
    \institution{Cambridge Quantum / Quantinuum}
    \country{United Kingdom}
}
\email{razin.shaikh@cambridgequantum.com}

\author{Stefano Gogioso}
\authornotemark[1]
\orcid{0000-0001-7879-8145}
\affiliation{%
    \institution{Hashberg Ltd}
    \country{United Kingdom}
}
\email{quantum@hashberg.io}


\begin{abstract}
We introduce a novel compositional description of Feynman diagrams, with well-defined categorical semantics as morphisms in a dagger-compact category.
Our chosen setting is suitable for infinite-dimensional diagrammatic reasoning, generalising the ZX calculus and other algebraic gadgets familiar to the categorical quantum theory community.

The Feynman diagrams we define look very similar to their traditional counterparts, but are more general: instead of depicting scattering amplitude, they embody the linear maps from which the amplitudes themselves are computed, for any given initial and final particle states.
This shift in perspective reflects into a formal transition from the syntactic, graph-theoretic compositionality of traditional Feynman diagrams to a semantic, categorical-diagrammatic compositionality.

Because we work in a concrete categorical setting---powered by non-standard analysis---we are able to take direct advantage of complex additive structure in our description.
This makes it possible to derive a particularly compelling characterisation for the sequential composition of categorical Feynman diagrams, which automatically results in the superposition of all possible graph-theoretic combinations of the individual diagrams themselves (cf. Figure \ref{fig:composition_calculation} above).

\end{abstract}



\keywords{Feynman diagrams, quantum field theory, categorical semantics, diagrammatic reasoning, formal diagrammatic methods}

\begin{teaserfigure}
  \begin{center}
  \includegraphics[width=0.96\textwidth]{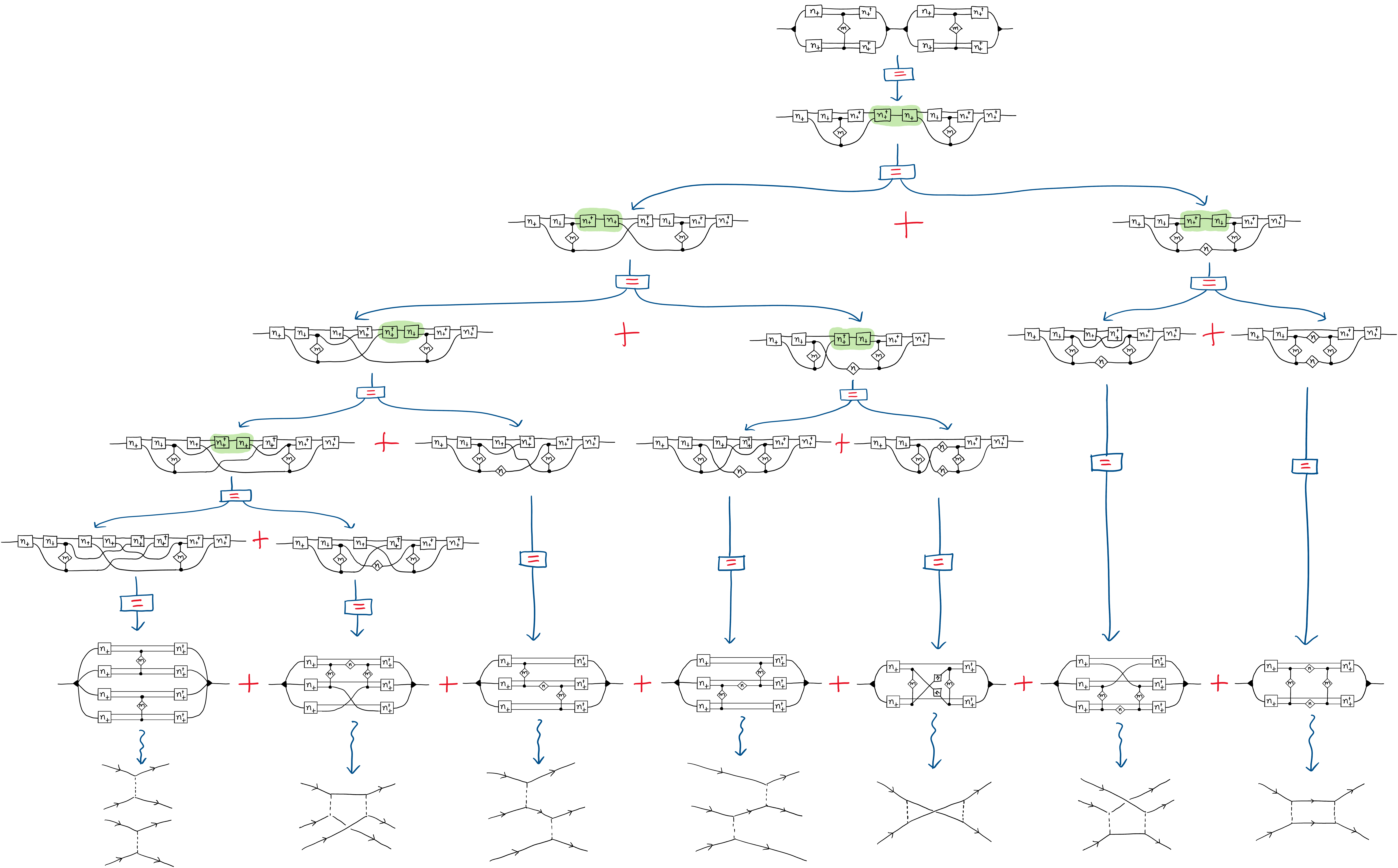}
  \caption{Composition of categorical Feynman diagrams, cf. Section~\ref{section:composing_feynman_diagrams}.}
  \Description{Sequential composition of categorical Feynman diagrams results in the superposition of all graph-theoretic compositions of the individual diagrams involved.}
  \label{fig:composition_calculation}
  \end{center}
\end{teaserfigure}

\maketitle

\section{Introduction}

Over the past decade, formal diagrammatic methods have evolved from novel foundational approach \cite{Abramsky2004CQM,Coecke2015CausalProcesses,Backens2014ZX,Coecke2011InteractingQuantum,Coecke2017PQP,Duncan2009GraphStates,Gogioso2015Schrodinger,Gogioso2017Thesis,Gogioso2018CPT,Heunen2019introCQM,Selinger2010SurveyGraphical} to established tools for quantum software design and compilation \cite{Coecke2021Kindergarden,Chancellor2018ErrorCorrection,deBeaudrap2020ReduceParityPhase,Duncan2020CircuitSimplification,Duncan2014SteaneCode,Duncan2010MBQC,Kissinger2019UniversalMBQC,Kissinger2020NonClifford,vandewetering2020zxcalculus}, with a significant portion of research progressively shifting from academia to industry.
As development continues apace, it seems likely---to the authors, at the very least---that these tools will ultimately catch up and overtake the more traditional formalisms that are prevalent at this time, establishing a new software development standard for the burgeoning field of digital quantum computation.

Axiomatic diagrammatic calculi---such as the ZX calculus~\cite{Coecke2011InteractingQuantum}, ZW calculus \cite{Hadzihasanovic2017thesis} or ZH calculus \cite{Backens2019ZH}---have thus far contributed the most to quantum compilation and optimisation, but more general diagrammatic methods based on monoidal category theory have further enriched our understanding of quantum information, as well as a variety of foundational topics in finite-dimensional quantum theory \cite{Brukner2014Causality,Chiribella2014OPT,Chiribella2010ProbabilisticTheories,Chiribella2019Shannon,Coecke2015CausalProcesses,Gogioso2019Dynamics,Kissinger2017CausalStructure,Pinzani2019TimeTravel,Pinzani2020OperationalSuperposition,Paunkovic2020CausalOrders}.
However, a long-standing limitation of such method has been their inability to rigorously describe most infinite-dimensional scenarios of concrete interest, from the wave-functions of textbook quantum mechanics all the way up to quantum field theory.
This has prevented diagrammatic methods from getting adoption in foundational and particle physics, quantum hardware design and branches of analog quantum computation---such as annealing and photonics---where quantum fields play a key role.
Our main contribution is to finally demonstrate the applicability of categorical diagrammatic methods to the aforementioned fields, by tackling field quantisation and incorporating Feynman diagrams.

The authors are only aware of a handful of attempts at infinite-dimensional extensions of such methods: a seminal work on $H^*$-algebras \cite{Abramsky2012HStar}, few works on Hopf algebras \cite{Collins2019,Majid2021} and a series of works using non-standard analysis (NSA) \cite{Gogioso2017InfiniteCQM,Gogioso2018TowardsQFT:,Gogioso2019QFTinCQM,Gogioso2019Dynamics,Robinson1974NonStandard}.
With the exception of perhaps one \cite{Gogioso2019QFTinCQM}, none of these attempts have come close to building the tools necessary to reason about quantum fields in a rigorous diagrammatic way.
Over years, the issue encountered by those who worked on such extensions has often been the inevitable emergence of infinities---infinities which physicists have long ago learned to safely handle, but which to date present a challenge for rigorous formalisation.
Other significant formalisation approaches to Feynman diagrams have also been pursued, most notably \cite{Blute2010ProofNets}, but from the viewpoint of logic rather than monoidal category theory.

The need for a rigorous handling of such infinities was the main motivation behind the non-standard approach to infinite-dimensional categorical quantum mechanics (CQM), a sequence of works of which the present paper is the latest instalment.
Non-standard analysis---one of the great 20th century triumphs of model theory---provides a way to work with infinite and infinitesimal quantities in an algebraic setting, without the need to work with limits or topological notions of convergence.
Calculations in the non-standard setting are performed in exact formal arithmetic, with the Transfer Theorem acting as the gateway through which results---such as the existence of interesting quantum states and maps---are transferred to and from the standard mathematical world.
In the aforementioned approach to infinite-dimensional CQM, NSA is used to define $\starHilbCategory$, a dagger compact category suitable for diagrammatic reasoning and able to accommodate the objects and morphisms necessary to talk about wave-functions and quantum fields.

On its own, the category $\starHilbCategory$ is rather simple: it contains hyperfinite non-standard complex Hilbert spaces and $\starComplexs$-linear maps between them. Hyper-finiteness is the key part of this definition: it means that the Hilbert spaces involved have dimension given by some non-standard natural number (i.e. they take the form $\starComplexs^d$ for some $d \in \starNaturals$), so by Transfer Theorem they possess many of properties of finite-dimensional Hilbert spaces.
In particular: $\starHilbCategory$ is a dagger compact category because so is $\fHilbCategory$;  it possesses special commutative dagger-Frobenius algebras ($\dagger$-SCFAs) because so does $\fHilbCategory$; we can form group algebras in $\starHilbCategory$---that is, strongly complementary pairs of $\dagger$-SCFAs---because we can do so in $\fHilbCategory$.
This last point turns out to be central to our endeavours: it allows the definition of quantised position and momentum spaces, connects them by Pontryagin duality and embodies the Weyl form of the Canonical Commutation Relations.

Thus said, the category $\starHilbCategory$ is only part of the story: it provides a rich canvas for rigorous diagrammatic reasoning, but its objects don't relate to any specific standard Hilbert space.
In fact, most constructions don't rely on a specific choice of dimension $d \in \starNaturals$ for the non-standard Hilbert space $\starComplexs^d$, with many or even all choices with $d$ infinite typically fitting the bill.
The physical meaning for each non-standard construction arises from an explicit map down to the standard world, typically involving restriction to \emph{near-standard elements} and application of the \emph{standard part function}.
Both concepts are \emph{external} in NSA parlance, which means that they cannot be reasoned about using the Transfer Theorem\footnote{To be technical, external sets and functions are those which don't appear in the relational structure for the non-standard model. The existence of such objects is key to non-standard set theory---also known as ``internal'' set theory---and is made possible by the use of non-full structures.}, but nonetheless they can be used to connect non-standard constructions to standard ones (taking extra care to mind one's steps).

The remainder of this work is structured as follows. Sections \ref{section:wavefunctions} and \ref{section:categorical_fields} summarise the necessary background from the non-standard approach to infinite-dimensional categorical quantum mechanics, including significant reformulation of existing material and some novel contributions about the Feynman propagator and the interaction picture. Section \ref{section:main_contribution} presents our main contributions: decompositional structure of the space of quantum fields, categorical Feynman diagrams, amplitude calculation and diagram composition.

\newpage
\section{Categorical wave-functions}
\label{section:wavefunctions}

\subsection{Non-standard lattice}

Consider a non-standard lattice in the following form:
\begin{equation}
    \frac{1}{\omega_{uv}} \starIntegersModPow{2\omega+1}{n}
    \defeq
    \setDefinition{
        \vec{x} \in \starReals^n
    }{
        \vec{x}=\frac{\vec{j}}{\omega_{uv}}, \ \vec{j} \in \starIntegersModPow{2\omega+1}{n}
    }
    \subset
    \starReals^n
\end{equation}
where $\omega_{uv}, \omega_{ir} \in \starNaturals$ are two odd non-standard natural numbers and $\omega \in \starNaturals$ is defined by $2\omega+1 = \omega_{uv}\omega_{ir}$.
The Abelian group $\starIntegersMod{2\omega+1}$ is defined by equipping the non-standard integers $\{-\omega, \dots, +\omega \}$ with addition modulo $2\omega+1$. The lattice inherits this group structure by rescaling, structure which depends on the choice of $\omega$ but not on the specific choice of $\omega_{uv}, \omega_{ir}$ such that $2\omega+1 = \omega_{uv}\omega_{ir}$ (were multiple choices to exist).
Using parlance from quantum field theory, we refer to $\omega_{uv}$ as the \emph{ultraviolet cut-off} and to $\omega_{ir}$ as the \emph{infrared cut-off}: the former determines the distance separating two points in the lattice---that is, the lattice \emph{mesh}---while the latter determines the overall volume $\omega_{ir}^n = \left((2\omega+1)/\omega_{uv}\right)^n$ of $\starReals^n$ spanned by the lattice.

\subsection{Standard quotient group}

Now we restrict our attention to the near-standard elements, i.e. those $\vec{x}$ which are infinitesimally close to a standard real number (necessarily unique and called $\stdpart{\vec{x}}$):
\begin{equation}
    \left(\frac{1}{\omega_{uv}} \starIntegersModPow{2\omega+1}{n}\right)_{fin}
    \hspace{-0.2mm}\defeq
    \setDefinition{
        \vec{x} \in \frac{1}{\omega_{uv}} \starIntegersModPow{2\omega+1}{n}
    }{
        \vec{x} \text{ near-standard}
    }
\end{equation}
Near-standard elements form an (external) sub-group, and taking the standard part defines a group homomorphism:
\begin{equation}
\begin{array}{rrcl}
    \stdpartSym:
    &\left(\frac{1}{\omega_{uv}} \starIntegersModPow{2\omega+1}{n}\right)_{fin}
    &\longrightarrow
    &\reals
    \\
    &\vec{x}
    &\mapsto
    &\operatorname{std}(\vec{x})
\end{array}
\end{equation}
We thus obtain a quotient group:
\begin{equation}
    G := \stdpart{\left(\frac{1}{\omega_{uv}} \starIntegersModPow{2\omega+1}{n}\right)_{fin}}
\end{equation}
Different quotient groups arise depending on whether none, one or both of $\omega_{uv}$ and $\omega_{ir}$ are infinite:
\begin{itemize}
    \item if $\omega_{uv}$ is finite and $\omega_{ir}$ is finite, we obtain finite abelian groups $G = \frac{1}{\stdpart{\omega_{uv}}}\integersMod{2\omega+1}^n \cong \integersMod{2\omega+1}^n$;
    \item if $\omega_{uv}$ is finite and $\omega_{ir}$ is infinite, we obtain the infinite group $G = \frac{1}{\stdpart{\omega_{uv}}}\integers^n \cong \integers^n$;
    \item if $\omega_{uv}$ is infinite and $\omega_{ir}$ is finite, we obtain the compact Lie group $G = \left(\reals/\left(\stdpart{\omega_{ir}}\integers\right)\right)^n \cong \mathbb{T}^n$;
    \item if $\omega_{uv}$ is infinite and $\omega_{ir}$ is infinite, we obtain the locally compact Lie group $G = \reals^n$.
\end{itemize}

\subsection{Dual non-standard lattice}

The dual lattice for $\frac{1}{\omega_{uv}} \starIntegersMod{2\omega+1}$ is obtained by swapping $\omega_{uv}$ (the small scale cut-off) and $\omega_{ir}$ (the large scale cut-off):
\begin{equation}
    \frac{1}{\omega_{ir}} \starIntegersModPow{2\omega+1}{n}
    =
    \setDefinition{
        \vec{p} \in \starReals^n
    }{
        \vec{p}=\frac{\vec{k}}{\omega_{ir}}, \ \vec{k} \in \starIntegersModPow{2\omega+1}{n}
    }
    \subset
    \starReals^n
\end{equation}
Restricting to near-standard elements of the dual lattice and taking the standard part yields the Pontryagin dual of $G$:
\begin{equation}
    G^\wedge
    =
    \stdpart{\left(\frac{1}{\omega_{ir}} \starIntegersModPow{2\omega+1}{n}\right)_{fin}}
\end{equation}
In practice, we use $\frac{1}{\omega_{uv}} \starIntegersModPow{2\omega+1}{n}$ to model positions in certain $n$-dimensional spaces of quantum-mechanical interest:
the associated group structure is that of translations on position space, the dual lattice $\frac{1}{\omega_{uv}} \starIntegersModPow{2\omega+1}{n}$ models the corresponding momentum space, and its group structure is that of momentum boosts (the Pontryagin dual of translations).

\subsection{Wave-functions on position space}

Given the above, the wave-functions on position space are modelled by the following object of $\starHilbCategory$:
\begin{equation}
    \starComplexs\left[\frac{1}{\omega_{uv}} \starIntegersMod{2\omega+1}\right]
\end{equation}
Within this Hilbert space, we can restrict our attention to those near-standard vectors $\ket{\psi}$ such that the inner product $\braket{\psi|\psi}$ is itself near-standard.
Taking $\stdpart{\ket{\psi}}$, we obtain exactly the functions of $L^2\left[G\right]$, that is we recover the standard picture of wave-functions on position space.
For this correspondence to hold, the inner product must correctly account for the volume element $d^n\vec{x} = 1/\omega_{uv}^n$:
\begin{equation}
    \braket{\psi|\varphi}
    \defeq
    \hspace{-3mm}
    \sum_{\vec{x} \in \frac{1}{\omega_{uv}} \starIntegersModPow{2\omega+1}{n}}
    \hspace{-3mm}
    \frac{1}{\omega_{uv}^n}
    \psi^*(\vec{x}) \varphi(\vec{x})
\end{equation}
There also exist finite vectors---and, in particular, normalised ones---which are not near-standard: these are genuinely new objects made available by the non-standard framework.
Importantly, these objects can be used to rigorously model notoriously tricky mathematical notions used in everyday quantum physics calculations, such as delta functions (which are now genuine vectors) and plane-waves (which can now be normalised).
Delta functions and plane-waves are, in fact, related to a familiar object from categorical diagrammatics: a strongly complementary pair $(\greenSpider, \redSpider)$ of quasi-special commutative $\dagger$-Frobenius algebras.
Specifically, we can take $\greenSpider$ to be the quasi-special commutative $\dagger$-Frobenius algebras copying the basis of delta functions, or \emph{position eigenstates}:
\begin{equation}
    \ket{\delta_{\vec{x}}}
    :=
    \vec{y} \mapsto
    \begin{cases}
    \omega_{uv}^n &\text{ if } \vec{x} = \vec{y}\\
    0 &\text{ otherwise}
    \end{cases}
\end{equation}
We then define a quasi-special commutative $\dagger$-Frobenius algebra $\redSpider$ by taking its multiplication and unit to define the $\frac{1}{\omega_{uv}} \starIntegersModPow{2\omega+1}{n}$ group structure on delta functions, yielding a strongly complementary pair $(\greenSpider, \redSpider)$.
The states copied by $\redSpider$ are the plane-waves, or \emph{momentum eigenstates}, indexed by all momentum values $\underline{p} \in \frac{1}{\omega_{uv}} \starIntegersModPow{2\omega+1}{n}$:
\begin{align}
    \ket{\goodchi_{\vec{p}}}
    &:=
    \hspace{-3mm}
    \sum_{\vec{x} \in \frac{1}{\omega_{uv}} \starIntegersModPow{2\omega+1}{n}}
    \hspace{-3mm}
    \frac{1}{\omega_{uv}^n}
    e^{i2\pi \vec{p} \cdot \vec{x}}
    \ket{\delta_{\vec{x}}}
    \nonumber \\
    &= \vec{x} \mapsto e^{i2\pi \vec{p} \cdot \vec{x}}
\end{align}
Neither position nor momentum eigenstates are normalised, their norms instead reflecting the volume elements of position and momentum space respectively:
\begin{equation}
    \braket{\delta_{\vec{x}}|\delta_{\vec{x}}}
    =\omega_{uv}^n
    \hspace{1cm}
    \braket{\goodchi_{\vec{p}}|\goodchi_{\vec{p}}}
    =\omega_{ir}^n
    \hspace{1cm}
    \braket{\delta_{\vec{x}}|\goodchi_{\vec{p}}}
    =e^{i2\pi \vec{p} \cdot \vec{x}}
\end{equation}

\section{Categorical quantum fields}
\label{section:categorical_fields}

\subsection{Field quantisation}

Consider the relativistic real scalar field whose Lagrangian is given by:
\begin{equation}
    \frac{1}{2} \p_{\mu} \phi \p^\mu \phi - \frac{1}{2} m^2 \phi^2
\end{equation}
This field gives rise to the Klein-Gordon equation:
\begin{equation}
    \p_\mu \p^\mu \phi + \left(\frac{mc}{\hbar}\right)^2 \phi = 0
\end{equation}
In momentum space, the Klein-Gordon equation factorises:
\begin{equation}
    \left(\hbar^2 \frac{\p^2}{\p t^2} + (|\vec{p}|c)^2 + (mc^2)^2\right) \phi(\vec{p}, t) = 0
\end{equation}
This is the equation of motion for a field of independent simple harmonic oscillators at each point $\vec{p} \in \reals^3$ of momentum space.
The oscillator at point $\vec{p}$ oscillates at a frequency $\nu_{\vec{p}} = \frac{1}{\hbar}E_{\vec{p}}$, where $E_{\vec{p}}$ is the energy given by the following \emph{relativistic dispersion relation}:
\begin{equation}
    E_{\vec{p}} \defeq \sqrt{(|\vec{p}|c)^2 + (mc^2)^2}
\end{equation}
We now proceed to quantise this field in $\starHilbCategory$.

Traditionally, $\ell^2(\naturals; \complexs)$ is considered as the space for quantised oscillators.
Its non-standard counterpart $\ell^2(\starNaturals; \starComplexs)$ is, however, infinite-dimensional, and $\starHilbCategory$ is limited (so to speak) to hyperfinite-dimensional non-standard Hilbert spaces.
To achieve our quantisation in $\starHilbCategory$, we consider versions of $\ell^2(\starNaturals; \starComplexs)$ which are ``truncated'' at some infinite particle number $\kappa$:
\begin{equation}
    \SpaceH^{\kappa}
    \defeq
    \starComplexs^{\kappa}
\end{equation}
Now, write $\left(|n\rangle\right)_{n=0}^{\kappa-1}$ for the canonical orthonormal basis of $\starComplexs^{\kappa}$.
In order to work around our (necessary) technical modification to the quantised oscillator space, we define the ladder operators to connect spaces truncated at different particle numbers:
\begin{align}
    a_{\kappa}^{\dagger}:\SpaceH^{\kappa} &\to \SpaceH^{\kappa + 1}
    & a_{\kappa}:\SpaceH^{\kappa} &\to \SpaceH^{\kappa - 1} \nonumber \\
    \ket{n} &\mapsto \sqrt{n+1} \ket{n+1}
    & \ket{n} &\mapsto \sqrt{n} \ket{n-1}
\end{align}
where the special case $\sqrt{0} \ket{-1}$ for $n=0$ on the right is set to the zero vector of the Hilbert space by convention.
The ladder operators satisfy the usual canonical commutation relation:
\begin{equation}
    [a,a^{\dagger}] = id
\end{equation}
Under the hood, ladder operators at different truncations are used to make dimensions match: 
\begin{equation}
    a_{\kappa+1} a^\dagger_{\kappa} - a^\dagger_{\kappa-1} a_{\kappa} = id_{\kappa}
\end{equation}
Our 3-momentum space is modelled by the following non-standard lattice, in the case where $\omega_{uv}, \omega_{ir} \in \starNaturals$ are both odd infinite naturals:
\begin{equation}
    \Omega := \frac{1}{\omega_{ir}} \starIntegersModPow{2\omega+1}{3}
\end{equation}
Our 3-position space is modelled by the dual lattice:
\begin{equation}
    \SpaceX = \frac{1}{\omega_{uv}} \starIntegersModPow{2\omega+1}{3}
\end{equation}
Just as the quantisation $\SpaceH^\kappa$ for each individual oscillator is truncated at some arbitrary particle number $\kappa$, so the overall field quantisation will be truncated at some arbitrary function $\tau: \Omega \rightarrow \starNaturals$ selecting an arbitrary particle number truncation $\tau_{\vec{p}}$ \emph{for each momentum value} $\vec{p}$.
Putting everything together, the following tensor product of quantised oscillators will be our quantised field space:
\begin{equation}
    \SpaceH^{(\tau)}
    \defeq
    \bigotimes_{\vec{p} \in \Omega} \SpaceH^{\tau_{\vec{p}}}
\end{equation}
Its canonical product basis takes the following form, for all possible assignments ${\vec{n} \in \prod\limits_{\vec{p} \in \Omega} \{0,\dots,\tau_{\vec{p}}\}}$ of particle numbers to individual momentum values:
\begin{equation}
    \ket{\beta_{\vec{n}}}
    :=
    \bigotimes_{\vec{p} \in \Omega}\ket{n_{\vec{p}}}
\end{equation}

\subsection{Ladder operators}

For notational convenience, we define the following indicator functions for all momenta $\vec{p} \in \Omega$:
\begin{align}\label{eq:delta_p_function}
    \delta_{\vec{p}}: \Omega & \to \{0,1\} \nonumber \\ 
    \vec{q} &\mapsto \delta_{\vec{p}, \vec{q}}
\end{align}
where $\delta_{\vec{p}, \vec{q}}$ is 1 when $\vec{p}=\vec{q}$ and 0 otherwise.
For each momentum value $\vec{p} \in \Omega$, we define \emph{ladder operators}, acting non-trivially on the quantised harmonic oscillator factor for momentum value $\vec{p}$ only:
\begin{align}
    \begin{array}{rrcl}
      a_\tau^{\dagger}(\vec{p}):
    & \SpaceH^{(\tau)}
    & \longrightarrow
    & \SpaceH^{(\tau + \delta_{\vec{p}})}
    \\
    & \ket{\beta_{\vec{n}}}
    & \mapsto
    & \sqrt{\omega_{ir}^3} \sqrt{n_{\vec{p}}+1} \ket{\beta_{\vec{n} + \delta_{\vec{p}}}}
    \\\\
      a_\tau(\vec{p}):
    & \SpaceH^{(\tau)}
    & \longrightarrow
    & \SpaceH^{(\tau - \delta_{\vec{p}})}
    \\
    & \ket{\beta_{\vec{n}}}
    & \mapsto
    & \sqrt{\omega_{ir}^3} \sqrt{n_{\vec{p}}} \ket{\beta_{\vec{n} - \delta_{\vec{p}}}}
    \end{array}
\end{align}
By convention, we set $\ket{\beta_{\vec{n}}}$ to be the zero vector if there exists $\vec{p} \in \Omega$ such that $n_{\vec{p}} < 0$.
Similarly, we set $\SpaceH^{(\tau)} \defeq \starComplexs^0$ if there exists $\vec{p} \in \Omega$ such that $\tau_{\vec{p}} < 0$.
The operator $a^{\dagger}(\vec{p})$ is also known as a \emph{creation operator}, while the operator $a(\vec{p})$ is known as an \emph{annihilation operator}.

In the expressions above, the $\omega_{ir}^3$ factor above comes from the Euclidean volume element for our discretised 3-momentum space:
\begin{equation}
    \int_{\reals^3} d^3\vec{p}
    \longleftrightarrow
    \sum_{\vec{p} \in \Omega} \frac{1}{\omega_{ir}^3}
\end{equation} 
In particular, the Euclidean \emph{delta function} on discretised 3-momentum space takes the following form:
\begin{equation}
    \delta(\vec{p}-\vec{q}) := \omega_{ir}^3 \delta_{\vec{p}, \vec{q}}
\end{equation}
The above observation helps us make sense of the commutator for the relativistically normalised ladder operators:
\begin{equation}
\begin{aligned}
    \left[a(\vec{p}), a^{\dagger}(\vec{q}) \right]
    &=
    \delta(\vec{p}-\vec{q})\, id\\
    \left[a(\vec{p}), a(\vec{q})\right] &= 0 =
    \left[a^{\dagger}(\vec{p}), a^{\dagger}(\vec{q})\right]
\end{aligned}
\end{equation}
Under the hood, ladder operators at different truncations are used to make dimensions match once again:
\begin{equation}
    a_{\tau+\delta_{\vec{p}}}(\vec{p}) a^\dagger_{\tau}(\vec{p})
    -
    a^\dagger_{\tau-\delta_{\vec{p}}}(\vec{p}) a_{\tau}(\vec{p})
    =
    \omega_{ir}^3\, id_{\tau}
\end{equation}

\subsection{Particle states}

For conceptual and notational convenience, we also define the following orthogonal basis of \emph{particle states} (in momentum space), obtained from the \emph{free field vacuum} $\ket{\underline{0}} := \ket{\beta_{\vec{0}}}$ by any hyperfinite number of applications of the creation operators:
\begin{align}
    \ket{\vec{n}}
    &\defeq
    \bigotimes_{\vec{p} \in \Omega}
    \sqrt{2 E_{\vec{p}} \omega_{ir}^3}^{\,n_{\vec{p}}}
    \sqrt{n_{\vec{p}}!}\,
    \ket{n_{\vec{p}}}
    \nonumber \\
    &=
    \prod_{\vec{p} \in \Omega}
    \left(\sqrt{2 E_{\vec{p}}} a^\dagger(\vec{p})\right)^{n_{\vec{p}}}
    \ket{\underline{0}}
\end{align}
In other words, the particle state $\ket{\vec{n}}$ is the one containing exactly $n_{\vec{p}}$ particles with momentum $\vec{p}$ for each $\vec{p} \in \Omega$, together with an additional \emph{relativistic normalisation} factor $\sqrt{2 E_{\vec{p}}}$.
In particular, the 1-particle states for each momentum value $\vec{p} \in \Omega$ take the following simple form:
\begin{equation}
    \ket{\delta_{\vec{p}}}
    =
    \sqrt{2 E_{\vec{p}}}a^{\dagger}(\vec{p}) \ket{\underline{0}}
\end{equation}
where the particle number $\delta_{\vec{p}}$ is 1 at $\vec{p}$ and 0 everywhere else.
The relativistic normalisation factor $\sqrt{2 E_{\vec{p}}}$ comes from the Lorentz-invariant volume element for on-shell 4-momentum space (in its equivalent formulation for 3-momentum space):
\begin{equation}
    \int_{\reals^4} d^4p \delta(p_0-E_{\vec{p}})|_{p_0>0}
    =
    \int_{\reals^3} d^3\vec{p} \frac{1}{2 E_{\vec{p}}}
    \longleftrightarrow
    \sum_{\vec{p} \in \Omega} \frac{1}{\omega_{ir}^3} \frac{1}{2 E_{\vec{p}}}
\end{equation} 
In particular, the \emph{Lorentz-invariant delta function} on discretised 3-momentum space appears as the inner product between 1-particle states:
\begin{equation}
\braket{\delta_{\vec{p}}|\delta_{\vec{q}}} 
=2E_{\vec{p}}\delta(\vec{p}-\vec{q})   
\end{equation}

\subsection{Field operator}

In traditional quantum field theory formulations, the position-space field operator is obtained by Fourier transform of the ladder operators.
Translating the Lorentz-invariant integral into a hyperfinite sums, our position-space field operator takes the following form:
\begin{align}
    \phi(\vec{x})
    \defeq
    \sum_{\vec{p} \in \Omega} \frac{1}{\omega_{ir}^3} \frac{1}{\sqrt{2 E_{\vec{p}}}}
    \left[
        a(\vec{p}) e^{i 2\pi \vec{p} \cdot \vec{x}}
        +
        a^{\dagger}(\vec{p}) e^{-i 2\pi \vec{p} \cdot \vec{x}}
    \right]
\end{align}
Above, position $\vec{x}$ is taken in the discretised 3-position space $\Omega^\wedge$, the Pontryagin dual to our 3-momentum space $\Omega$.

Unlike traditional formulation of quantum field theory, we don't need to use field operators as a mathematical proxy for position-space field states. Indeed, the 1-particle state localised at a given position $\vec{x}$ can be written explicitly by Fourier transform, as a linear combination of 1-particle states in momentum space:
\begin{align}
    \ket{\delta_{\vec{x}}}
    &\defeq
    \phi(\vec{x}) \ket{\vec{0}}
    \nonumber \\
    &=
    \sum_{\vec{p} \in \Omega} \frac{1}{\omega_{ir}^3} \frac{1}{\sqrt{2 E_{\vec{p}}}}
    \left[
        a(\vec{p}) e^{i 2\pi \vec{p} \cdot \vec{x}}
        +
        a^{\dagger}(\vec{p}) e^{-i 2\pi \vec{p} \cdot \vec{x}}
    \right]\ket{\vec{0}}
    \nonumber \\
    &=
    \sum_{\vec{p} \in \Omega} \frac{1}{\omega_{ir}^3} \frac{1}{2 E_{\vec{p}}}
    e^{-i 2\pi \vec{p} \cdot \vec{x}} \ket{\delta_{\vec{p}}}
\end{align}

\subsection{Coherently-controlled ladder operators}

Consider the space of wave-functions on 3-position space:
\begin{equation}
    \starComplexs\left[\SpaceX\right]
\end{equation}
Additionally to the position eigenstates $\ket{\delta_{\vec{x}}}$ and momentum eigenstates $\ket{\goodchi_{\vec{p}}}$ defined in Section \ref{section:wavefunctions}, the following \emph{relativistically normalised} momentum eigenstates will also come in handy (note the slight shift in notation from 3-momentum $\vec{p}$ to 4-momentum $p$):
\begin{equation}\label{eq:rel_momentum_eigenstates}
    \ket{\goodchi_{p}}
    \defeq
    \sqrt{2E_{\vec{p}}}\ket{\goodchi_{\vec{p}}}
\end{equation}
Using the basis of momentum eigenstates from Section \ref{section:wavefunctions}, we define \emph{coherently controlled ladder operators} diagrammatically:
\begin{equation}
\label{fig:coherent_creation}
    \tikzfigscale{0.7}{figures/coherent_creation}
\end{equation}
\begin{equation}
\label{fig:coherent_annihilation}
\hspace{-1mm}
    \tikzfigscale{0.7}{figures/coherent_annihilation}
\end{equation}
By acting on the relativistically normalised momentum eigenstates with the coherently controlled ladder operators, we recover the ladder operators:
\begin{equation}
    \tikzfigscale{0.7}{figures/coherent_creation_recover}
\end{equation}
\begin{equation}
    \tikzfigscale{0.7}{figures/coherent_annihilation_recover}
\end{equation}
The beauty of defining coherently controlled---as opposed to classically controlled---ladder operators is that we can plug superpositions of control basis states into the control system.
In particular, acting on the position eigenstates we obtain the positive frequency part $\phi^+(\vec{x})$ and the negative frequency part $\phi^-(\vec{x})$ of the field operator $\phi(\vec{x})$, diagrammatically:
\begin{align}
    \tikzfigscale{0.7}{figures/field_frequency}
\end{align}

\subsection{Unitary dynamics on 3-position space}

To incorporate dynamics, we discretize time using a 1-dimensional lattice $\SpaceT := \frac{1}{\omega_{uv}} \starIntegersMod{2\omega+1}$, analogous to the lattice $\SpaceX$ used to discretized 3-position space.
The action of time-translation on momentum eigenstates $\ket{\goodchi_{\vec{p}}}$ is captured by the following unitary representation $(U_t)_{t \in \SpaceT}$ acting on $\starComplexs[\SpaceT]$:
\begin{equation}
    U_t \defeq \sum_{\vec{p} \in \Omega} \frac{1}{\omega_{ir}^3} \frac{1}{2E_{\vec{p}}} e^{i 2 \pi E_{\vec{p}} t} \ket{\goodchi_p} \bra{\goodchi_p}
\end{equation}
In order for the above to be well-defined, however, we have to slightly (i.e. infinitesimally) adjust our definition of the energy levels, to make them elements of the dual lattice for $\SpaceT$:
\begin{equation}
    E_{\vec{p}} \defeq \frac{1}{\omega_{ir}} \floor*{\omega_{ir} \sqrt{|\vec{p}|^2 + m^2}}
\end{equation}
On the space $\starComplexs\left[\SpaceT\right]$, there is a strongly complementary pair $(\greenSpider, \redSpider)$ for time-energy duality, with time states $\ket{\delta_t}$ copied by $\greenSpider$, and $\redSpider$ defining the group action of time translations on them.
We define a coherent version of the unitary representation above, in the form of a unitary module for the bialgebra $(\greenSpider, \redSpider)$ on $\starComplexs[\SpaceT]$:
\begin{equation}
    \tikzfigscale{0.7}{figures/heisenberg_unitary}
\end{equation}

\subsection{The Heisenberg picture}

We now move from the Schr\"odinger picture to the Heisenberg picture, where time-dependence is incorporated into an operator $O$ as:
\begin{equation}
    e^{i \frac{2\pi}{h} H t} \; O \; e^{-i \frac{2\pi}{h} H t}
\end{equation}
where $H$ is the Hamiltonian:
\begin{equation}
    H
    \defeq
    \sum_{\vec{p} \in \Omega}
    \frac{1}{\omega_{ir}^3} E_{\vec{p}} a^{\dagger} (\vec{p}) a (\vec{p})
\end{equation}
In particular, we get the following ladder operators in the Heisenberg picture:
\begin{align}
    e^{i \frac{2\pi}{h} H t} a^{\dagger}(\vec{p}) e^{-i \frac{2\pi}{h} H t}
    &= e^{+iE_{\vec{p}}t} a^{\dagger}(\vec{p})\\
    e^{i \frac{2\pi}{h} H t} a(\vec{p}) e^{-i \frac{2\pi}{h} H t}
    &= e^{-iE_{\vec{p}}t} a(\vec{p})
\end{align}
where the equalities on the right follow from the commutation relations $[H, a^{\dagger}(\vec{p})] = +E_{\vec{p}} a^{\dagger}(\vec{p})$ and  $[H, a(\vec{p})] = -E_{\vec{p}} a(\vec{p})$.
It's thus enough to act on the parameter space to define a Heisenberg picture version of the coherently controlled ladder operators:
\begin{equation}
    \tikzfigscale{0.7}{figures/coherent_creation_heisenberg}
\end{equation}
We will drop the subscript $H$: we will not need the Schr\"odinger picture operators for the rest of this work.
From this, we get the Lorentz invariant version of the Field operator, where the notation $p \cdot x$ is defined to mean $E_{\vec{p}}t - \vec{p} \cdot \vec{x}$:
\begin{equation}
    \phi(\vec{x},t) \defeq \phi^+(\vec{x},t) + \phi^-(\vec{x},t)
\end{equation}
\begin{equation}
\label{eq:field_frequency_positive_heisenberg}
    \phi^+(\vec{x},t)
    :=\ 
    \tikzfigscale{0.7}{figures/field_frequency_positive_heisenberg}
    \ =\ 
     \sum_{\vec{p} \in \Omega} \frac{1}{\omega_{ir}^3} \frac{1}{\sqrt{2E_{\vec{p}}}} a(\vec{p}) e^{-i 2 \pi p \cdot x}    
\end{equation}
\begin{equation}
\label{eq:field_frequency_negative_heisenberg}
    \phi^-(\vec{x},t)
    :=\ 
    \tikzfigscale{0.7}{figures/field_frequency_negative_heisenberg}
    \ =\ 
    \sum_{\vec{p} \in \Omega} \frac{1}{\omega_{ir}^3} \frac{1}{\sqrt{2E_{\vec{p}}}} a^{\dagger}(\vec{p}) e^{+i 2 \pi p \cdot x} 
\end{equation}
From now on, we will group 3-position $\vec{x}$ and time $t$ together into a 4-position $x$, using the space of wave-functions over spacetime $\starComplexs[\SpaceX]\otimes\starComplexs[\SpaceT]$ as our control system. We will also group the two wires (and the associated Frobenius algebras) into a single one, for notational simplicity:
\begin{equation}
    \phi^+(x) \quad = \ \quad \tikzfigscale{0.7}{figures/field_frequency_positive_heisenberg_four}
\end{equation}
\begin{equation}
    \phi^-(x) \quad = \ \quad \tikzfigscale{0.7}{figures/field_frequency_negative_heisenberg_four}
\end{equation}
Furthermore, we will adopt $\ket{\delta_x}$ and $\ket{\goodchi_p}$ as notation for the position/momentum eigenvalues of this combined control system, where 4-momentum is $p=(\vec{p}, E_{\vec{p}})$.

\subsection{Complex scalar fields}

The machinery developed in this section straightforwardly generalises to the case of complex scalar fields, governed by the following Lagrangian:
\begin{equation}
    \lagr_{\text{complex}} = \p_{\mu} \bar{\psi} \p^{\mu} \psi - M^2 \bar{\psi}\psi
\end{equation}
This Lagrangian gives rise to two equations of motion:
\begin{align}
    \p_{\mu} \p^{\mu} \psi + M^2 \psi &= 0\\
    \p_{\mu} \p^{\mu} \bar{\psi} + M^2 \bar{\psi} &= 0
\end{align}
As such, the complex scalar field is comprised of two species of particle---a particle $\psi$ and the corresponding anti-particle $\bar{\psi}$---each with its own quantised field space:
\begin{equation}
    \SpaceH^{(\tau)} \otimes \SpaceH^{(\sigma)}
\end{equation}
The particle states for such a complex scalar field arise by tensor product.
Each particle specie has its pair of ladder operators, commuting with the operator of the other specie:
\begin{equation}
\begin{aligned}
    \left[b(\vec{p}), b^{\dagger}(\vec{q})\right] &= \delta(\vec{p}-\vec{q})\, id\\
    \left[c(\vec{p}), c^{\dagger}(\vec{q})\right] &= \delta(\vec{p}-\vec{q})\, id\\
    \left[b(\vec{p}), c^{\dagger}(\vec{q})\right] &= 0 = \left[b^\dagger(\vec{p}), c(\vec{q})\right]
\end{aligned}
\end{equation}
The propagator is now defined by commutation of the particle and antiparticle field operators, and the Feynman propagator is defined from it as in the real scalar case:
\begin{equation}
    D(x-y) = \left[\psi(x), \psi^{\dagger}(y) \right]
\end{equation}

\subsection{The Feynman propagator}

The \emph{propagator} $D(x-y)$ tells us the amplitude of a particle at spacetime point $x$ to reach the spacetime point $y$.
It is defined as:
\begin{equation} \label{eq:propagator_commutation}
    D(x-y)
    \defeq
    \left[\phi^{+} (x), \phi^{-} (y)\right]
    =
    \sum_{\vec{p} \in \Omega}\frac{1}{\omega_{ir}^3} \frac{1}{2E_{\vec{p}}} e^{-i 2 \pi p \cdot (x-y)} id
\end{equation}
The \emph{Feynman propagator} is defined as the time-ordered version of the propagator:
\begin{equation}
    \Delta_F (x-y) =
    \begin{cases}
        D(x-y) & x_0 > y_0\\
        D(y-x) & y_0 > x_0
    \end{cases}
\end{equation}
The Feynman propagator is not influenced by the order of parameters $x$ and $y$, so we have:
\begin{equation}
    \Delta_F (x-y) = \Delta_F (y-x)
\end{equation}
To reflect this property, we will introduce a symmetric notation for the Feynman propagator, where $\phi$ indicates the free field in question:
\begin{equation}
    \tikzfigscale{0.6}{figures/feynman_propagator}
\end{equation}
More specifically, the Feynman propagator is self-adjoint with respect to the 4-position $\greenSpider$ structure:
\begin{equation}
    \tikzfigscale{0.6}{figures/feynman_propagator_2}
\end{equation}

\subsection{The interaction picture}

In perturbative quantum field theory, it is useful to work in the \emph{interaction picture}, a hybrid representation where the Heisenberg picture is used for the free-field Hamiltonian $H_0$ and the Schr\"odinger picture is used for the interaction Hamiltonian $H_{\text{int}}$:
\begin{equation}
    H = H_0 + H_{\text{int}}
\end{equation}
More specifically, $H_0$ will govern the time dependence of operators while $H_{\text{int}}$ will govern the time dependence of states.
Typically, $H_0$ corresponds to a system of free fields and, as such, it is typically well-understood and exactly soluble.
In contrast, $H_{\text{int}}$ is a perturbation to the system that represents interaction between the fields, typically hard to handle non-perturbatively.
The states and operators are transformed from Schr\"odinger to interaction picture as follows:
\begin{align}
    \ket{\psi(t)}_I &= e^{iH_0t}\ket{\psi(t)}_S\\
    O_I(t) &= e^{iH_0t} O_S e^{-iH_0t}
\end{align}
where the subscripts $S$ and $I$ denote the Schr\"odinger and the interaction picture respectively.
The free field Hamiltonian $H_0$ in the interaction picture remains the same as in the Schr\"odinger picture:
\begin{equation}
    (H_0)_I = e^{iH_0t} (H_0)_S e^{-iH_0t} = (H_0)_S
\end{equation}
On the other hand---since $H_0$ does not typically commute with $H_{\text{int}}$---we get a different expression for the interaction Hamiltonian $H_{\text{int}}$ in the interaction picture, denoted by $H_I$:
\begin{equation}
    H_I \defeq (H_{\text{int}})_I = e^{iH_0t} (H_{\text{int}})_S e^{-iH_0t}
\end{equation}
From Schr\"odinger's equation in the Schr\"odinger picture, one can derive the Schr\"odinger equation in the interaction picture:
\begin{equation}\label{eq:int_schrodinger_eq}
\begin{aligned}
    i \frac{d}{dt}\left( \ket{\psi}_I \right) &= H_I(t) \ket{\psi}_I\\
    \ket{\psi(t)}_I &= U_{t_0, t} \ket{\psi(t_0)}_I
\end{aligned}
\end{equation}
Above, $\left( U_{t_0, t}\right)_{t_0, t \in \SpaceT} $ is a unitary representation responsible for time-translation on particle states, given by Dyson's Formula:
\begin{equation}\label{eq:dysons_formula}
    U_{t_0, t} = T\exp \left(-i \sum_{t'= t_0}^{t} \frac{1}{\omega_{uv}} H_I (t') \right) id
\end{equation}
The \emph{time ordering operator} $T$ rearranges the operators evaluated at later times to the left of operators evaluated at earlier times, e.g.
\begin{equation}
    T\{O_1(t_1) O_2(t_2)\} = \begin{cases}
        O_1(t_1) O_2(t_2) \qquad t_1 > t_2\\
        O_2(t_2) O_1(t_1) \qquad t_2 > t_1
    \end{cases}
\end{equation}
Moreover, note that the exponential in Dyson's formula is defined in terms of a Taylor series expansion:
\begin{equation}\label{eq:dyson_series}
\begin{aligned}
      &\exp \left(-i \sum_{t'= t_0}^{t} \frac{1}{\omega_{uv}} H_I (t') \right)\\
    = &1 - i \sum_{t'= t_0}^{t} \frac{1}{\omega_{uv}} H_I (t') + \frac{(-i)^2}{2} \left(\sum_{t'= t_0}^{t} \frac{1}{\omega_{uv}} H_I (t') \right)^2 + \dots
\end{aligned}
\end{equation}

\subsection{Scattering}
The process by which the incoming particles transition to the outgoing particles under the influence of an interaction is called \emph{scattering}.
In quantum field theory, one is therefore interested in computing the associated \emph{scattering amplitudes}, from which other quantities of physical interest---such as cross-sections and decay rates---can be derived.
To calculate amplitudes, we need to make a non-trivial assumption: that the initial state $\ket{i}$ is (approximately) free in far past $t=-\infty$ and that final state $\ket{f}$ is (approximately) free in the far future $t=+\infty$.
As our values for $\pm \infty$ we take:
\begin{equation}
    \pm \infty := \pm \frac{\omega}{\omega_{uv}} \approx \pm \frac{\omega_{ir}}{2}
\end{equation}
The scattering amplitude from $\ket{i}$ to $\ket{f}$ is $\Braket{f | U_{-\infty, +\infty} | i}$, where the unitary operator $U_{-\infty, +\infty}$ is itself known as the \emph{S-matrix}.
Feynman diagrams represent individual terms in the Taylor expansion of the S-matrix: if we compute the amplitudes associated to progressively higher order diagrams, we obtain increasingly precise approximations\footnote{For perturbative theories, that is.} for the scattering amplitudes, and all quantities that depend on them.

\section{Categorical Feynman diagrams}
\label{section:main_contribution}

In this section, we develop a compositional, categorical description of Feynman diagrams as processes in $\starHilbCategory$.
The starting point for this description is the connection between Feynman diagrams and terms in Wick's expansion for the time-ordered products of Dyson's Formula.
For a recap of the relevant background on Wick's Theorem and Feynman diagrams, please refer to Section~\ref{appendix:feynman_diagrams} in the Appendix.

To concretely exemplify our construction, we work in scalar Yukawa theory: a complex scalar field $\psi$ of \emph{nucleons} and \emph{anti-nucleons}, with ``strong force'' interactions mediated by a real scalar field $\phi$ of mesons.
We adopt the following notation for the associated coherently controlled ladder operators and Feynman propagators:
\begin{itemize}
    \item $n_+^{\dagger}$ and $n_+$ for the ladder operators of nucleons
    \item $n_-^{\dagger}$ and $n_-$ for the ladder operators of anti-nucleons
    \item $n$ for the Feynman propagator of nucleons and anti-nucleons
    \item $m^{\dagger}$ and $m$ for the ladder operators of mesons
    \item $m$ for the Feynman propagator of mesons
\end{itemize}
Consider the following 2nd order Feynman diagram for nucleon-nucleon scattering:
\begin{equation}
\label{eq:NNoder2_feynman_diagram}
    \vcenter{\hbox{\scalebox{0.7}{$
    \feynmandiagram [vertical=b to a] {
        i1 
        -- [anti fermion] a 
        -- [anti fermion] f1,
        
        a -- [scalar] b,
        
        i2 
        -- [anti fermion] b 
        -- [anti fermion] f2,
        
        f1 -- [draw=none] y -- [draw=none] f2,
        i1 -- [draw=none] x -- [draw=none] i2,
    };
    $}}}
\end{equation}
This diagram arises from the 2nd order term in the S-matrix of scalar Yukawa theory:
\begin{equation}
    \frac{(-ig)^2}{2} \sum_{x_1, x_2} \frac{1}{\omega_{uv}^8} 
    T \left\{\psi^{\dagger} (x_1) \psi(x_1) \phi(x_1) \psi^{\dagger} (x_2) \psi(x_2) \phi(x_2) \right\}
\end{equation}
Specifically, it corresponds to the following normally ordered term in the expansion of the above time-ordered term given by Wick's Theorem:
\begin{align}
    &\frac{(-ig)^2}{2} \sum_{x_1, x_2} \frac{1}{\omega_{uv}^8} 
    : n_+^{\dagger}(x_1) n_+(x_1)  n_+^{\dagger}(x_2) n_+(x_2) :
    \wick{\c {\phi(x_1)} \c{\phi(x_2)}}\\
    =\ & \frac{(-ig)^2}{2} \sum_{x_1, x_2} \frac{1}{\omega_{uv}^8} 
    n_+^{\dagger}(x_1) n_+^{\dagger}(x_2) n_+(x_1) n_+(x_2)
    \wick{\c {\phi(x_1)} \c{\phi(x_2)}} \label{eq:nucleon_scattering_ladder}
\end{align}
Translating this expression into a categorical diagram, we obtain a linearised version of the Feynman diagram as a process in $\starHilbCategory$:
\begin{equation} \label{eq:nucleon_scattering_ladder_not_2d}
    \tikzfigscale{0.55}{figures/nucleon_scattering_ladder_operators_inline}
\end{equation}
This might not look like a Feynman diagram---and certainly possesses none of their intuitive beauty---but it describes the correct process: all field operators in a normally ordered term act on the same space of quantum fields, in sequence, so the associated process is naturally 1-dimensional.
In order to understand the nature of its 2-dimensionality, we'll need to do some legwork.

\subsection{Split and merge maps}\label{sec:split_merge}

Key to our compositional, 2-dimensional interpretation of Feynman diagrams is the following observation: processes $a_1,...,a_k$ which act sequentially on a same space $\SpaceH$ but commute with each other, can alternatively be thought as acting in parallel on different ``portions'' of $\SpaceH$, as a tensor product.
The only ingredient necessary to formalise this intuition is maps that ``split'' a single copy of a space $\SpaceH$ into multiple copies that can be acted upon independently by the processes $a_1,...,a_k$, and then ``merge'' the result back into a single copy.
This will reveal Feynman diagrams as a naturally ``decompositional''---as opposed to compositional---processes: instead of arising by tensor product composition of simpler processes, their 2-dimensional description is actually the consequence of a decomposition of a single underlying space into non-interacting sectors.

In order to define suitable split and merge maps, we consider a particle state $\ket{\vec{n}}$---for some given choice of particle numbers $\vec{n} \in \prod_{\vec{p} \in \Omega} \{0,\dots, \goodtau(\vec{p})\}$---and define the set $\Theta_{\vec{n}}^{k}$ of all possible ways of partitioning the particles of $\ket{\vec{n}}$ into $k$ independent copies of $\SpaceH$:
\begin{equation}
    \Theta_{\vec{n}}^{k}
    \defeq
    \setDefinition{
        (\vec{i_1}, \dots, \vec{i_k})
        \in
        \left(\prod_{\vec{p} \in \Omega} \{0,\dots, \goodtau(\vec{p})\}\right)^k
    }{
        \vec{i_1} + \dots + \vec{i_k} = \vec{n}
    }
\end{equation}
The \emph{split maps} are morphisms defined as follows:
\begin{align}
    \tikzfigscale{0.7}{figures/split_map_noalpha} \ \ :&\, \SpaceH^{(\goodtau)} \rightarrow \, \underbrace{\SpaceH^{(\goodtau)} \, \otimes \, \dots \, \otimes \, \SpaceH^{(\goodtau)}}_{k}\\
    &\ket{{\vec{n}}} \mapsto \sqrt{\braket{{\vec{n}}|{\vec{n}}}} \sum_{\uuline{i} \in \Theta_{\vec{n}}^{k}} 
    \bigotimes_{j=1}^k \frac{1}{\sqrt{\braket{\vec{i_j}|\vec{i_j}}}}\ket{\vec{i_j}}
\end{align}
The \emph{merge maps} are morphisms defined as follows:
\begin{align}
    \tikzfigscale{0.7}{figures/merge_map_noalpha} \ \ :& \, \underbrace{\SpaceH^{(\goodtau)} \, \otimes \, \dots \, \otimes \, \SpaceH^{(\goodtau)}}_{k} \rightarrow \SpaceH^{(\goodtau)} \\
    &\bigotimes_{j=1}^k \ket{\vec{i_j}}  \mapsto \left(\prod_{j=1}^k \sqrt{\Braket{\vec{i_j}|\vec{i_j}}} \right)
    \frac{1}{\sqrt{\Braket{{\vec{n}} | {\vec{n}}}}}\ket{{\vec{n}}}
\end{align}
where above we have used the shorthand $\vec{n} := \vec{i_1} + \dots + \vec{i_k}$.
The split and merge maps for the same $k$ are adjoint, with split maps being isometries (a fact which can be readily verified):
\begin{align}\label{eq:split_merge_isometry}
    \tikzfigscale{0.7}{figures/split_merge_isometry_noalpha}
\end{align}
However, merge maps for $k>1$ cannot be isometries, because the hyperfinite dimensionality of their domain is $k$ times that of their codomain:
\begin{equation}
    \tikzfigscale{0.7}{figures/split_merge_not_unitary_noalpha}
\end{equation}

In order to obtain our 2-dimensional description of Feynman diagrams, we need to connect their fundamental sequential components---the ladder operators---to the split and merge maps we just defined.
This turns out to be rather straightforward, and hinges on the interpretation of creation and annihilation operators as, rather unimaginatively, creating and annihilating particles: see Section~\ref{appendix:sliding_rules} for a short explanation.
In practice, the relationship between split maps and annihilation operators is captured by the following equation, which we refer to as a \emph{sliding rule}:
\begin{equation}
    \tikzfigscale{0.6}{figures/sliding_rule_split}
\end{equation}
Taking the adjoint of the equation above yields a dual sliding rule for creation operators and merge maps:
\begin{equation}
    \tikzfigscale{0.6}{figures/sliding_rule_merge}
\end{equation}
Repeated application of the sliding rules---which are valid for all branches---yields derived sliding rules for sequences of operators:
\begin{equation}
    \tikzfigscale{0.6}{figures/sliding_rule_split_multiple}
\end{equation}
\begin{equation}
    \tikzfigscale{0.6}{figures/sliding_rule_merge_multiple}
\end{equation}

\subsection{Categorical Feynman diagrams}
\label{sec:categorical_feynaman_diagrams_derivation}

We have finally reached the point where our Feynman diagrams become 2-dimensional, courtesy of the decompositional machinery introduced in the previous section.
We start with the 1-dimensional presentation from before:
\begin{equation}
    \tikzfigscale{0.55}{figures/nucleon_scattering_ladder_operators_inline}
\end{equation}
We use a split-merge pair to decompose $\SpaceH$ into two independent copies:
\begin{equation}
    \tikzfigscale{0.55}{figures/nucleon_scattering_ladder_operators_isometry}
\end{equation}
We use the sliding rules to move the annihilation and creation operations onto the branches:
\begin{equation}
    \tikzfigscale{0.55}{figures/nucleon_scattering_ladder_operators_sliding}
\end{equation}
We replace summations over 4-positions with $\greenSpider$ spiders:
\begin{equation}
    \tikzfigscale{0.55}{figures/nucleon_scattering_ladder_operators_spiders}
\end{equation}
We fuse the spiders, pull all wires taut, and finally get:
\begin{equation}
    \tikzfigscale{0.55}{figures/nucleon_scattering_ladder_operators_feynman_diagram}
\end{equation}
This categorical diagram is functionally equivalent to the 2nd order Feynman diagram from Equation~\eqref{eq:NNoder2_feynman_diagram}, with the addition of split and merge maps making it clear that the 2-dimensional presentation is the effect of quantum field space decomposition.
Let's now consider the more sophisticated example of a loop diagram:
\begin{equation}
\label{eq:NNoder4_feynman_diagram}
\vcenter{\hbox{\scalebox{0.7}{$
\feynmandiagram [vertical=d to a] {
    i1 
    -- [fermion] a 
    -- [fermion] f1,
    
    a -- [scalar] b,
    b -- [plain, half left, looseness=1.75] c -- [anti fermion, half left, looseness=1.75] b,
    c -- [scalar] d,

    i2 
    -- [fermion] d
    -- [fermion] f2,
    
    i1 -- [draw=none] x1 -- [draw=none] x2 -- [draw=none] x3 -- [draw=none] i2,
    f1 -- [draw=none] y1 -- [draw=none] y2 -- [draw=none] y3 -- [draw=none] f2,
};
$}}} 
\end{equation}
The Wick's expansion term for this diagram is:
\begin{equation}
\scalebox{0.97}{$\displaystyle
    \frac{(-ig)^4}{4!} \frac{1}{(\omega_{uv}^4)^4} \sum_{x_1, x_2, x_3, x_4} 
    \left(
    \begin{aligned}
        n_+^{\dagger}(x_1) n_+^{\dagger}(x_2) n_+(x_1) n_+(x_2)\\
        \wick{\c1 {\phi(x_1)} \c2 {\phi(x_2)} \c1 {\phi(x_3)} \c2 {\phi(x_4)}}&\\
        \wick{\c3 {\psi^{\dagger} (x_3)} \c2 {\psi (x_3)} \c2 {\psi^{\dagger} (x_4)} \c3 {\psi (x_4)}}&
    \end{aligned}
    \right)
$}
\end{equation}
We write the term as a 1-dimensional categorical diagram:
\begin{equation}
    \tikzfigscale{0.51}{figures/nucleon_scattering_loop_inline}
\end{equation}
We insert a split-merge pair:
\begin{equation}
    \tikzfigscale{0.54}{nucleon_scattering_loop_isometry}
\end{equation}
We use the sliding rules to rearrange ladder operators onto the branches:
\begin{equation}
    \tikzfigscale{0.55}{nucleon_scattering_loop_sliding}
\end{equation}
We replace summations with spiders:
\begin{equation}
    \tikzfigscale{0.55}{nucleon_scattering_loop_spiders}
\end{equation}
We fuse spiders and pull wires taut:
\begin{equation}
    \tikzfigscale{0.55}{nucleon_scattering_loop_feynman_diagram}
\end{equation}
This categorical diagram is functionally equivalent to the 4th order loop Feynman diagram from Equation~\eqref{eq:NNoder4_feynman_diagram}.
Further worked-out examples of interest can be found in Section~\ref{appendix:more_categorical_feynman_diagrams} of the Appendix.

\subsection{Computing amplitudes}
\label{sec:recovering_amplitudes}

Computation of amplitudes is simply a matter of plugging in the desired initial state and testing against the desired final co-state.
We show here a condensed example: for the full calculations, see Section~\ref{appendix:computing_amplitudes} of the Appendix.

Let's go back to our nucleon-nucleon scattering example:
\begin{equation}
    \tikzfigscale{0.5}{figures/nucleon_scattering_ladder_operators_feynman_diagram}
\end{equation}
We wish to compute the amplitude for the following initial and final states:
\begin{align*}
    \ket{i} &= \ket{p_1, p_2} = n_+^{\dagger}(p_1) n_+^{\dagger}(p_2) \ket{0}\\
    \ket{f} &= \ket{p'_1, p'_2} = n_+^{\dagger}(p'_1) n_+^{\dagger}(p'_2) \ket{0}
\end{align*}
We plug these initial and final states in the above categorical diagram (omitting the factor $\frac{(-ig)^2}{2} \frac{1}{\omega_{uv}^8}$ for convenience):
\begin{equation}
\begin{aligned}
    &\quad \tikzfigscale{0.55}{figures/feynman_amplitude_1}\\[1em]
    =&\quad \tikzfigscale{0.55}{figures/feynman_amplitude_3}\label{eq:feynman_amplitude_3}
\end{aligned}    
\end{equation}
To simplify this last scalar diagram, we will use the commutation relation:
\begin{equation}
\label{eq:nucleon_commutation_relation}
    \tikzfigscale{0.55}{figures/nucleon_commutation_relation}
\end{equation}
Specifically, we will apply it to the vacuum state:
\begin{equation}
    \tikzfigscale{0.55}{figures/nucleon_commutation_relation_vacuum}
\end{equation}
Using the commutation relation a few times, we get:
\begin{align}
    &\quad \tikzfigscale{0.55}{figures/feynman_amplitude_4} \nonumber\\[1em]
    =&\quad \tikzfigscale{0.55}{figures/feynman_amplitude_6}
\end{align}
We perform a symmetric simplification on the final co-state, use the fact that the vacuum state is normalized and reintroduce the $\frac{(-ig)^2}{2} \frac{1}{\omega_{uv}^8}$ factor:
\begin{equation}\label{eq:feynman_amplitude_diagram}
    (-ig)^2 \frac{1}{\omega_{uv}^8} \left( \  \tikzfigscale{0.55}{figures/feynman_amplitude_8} \ \right)
\end{equation}
The above is functionally equivalent to the corresponding sum of traditional Feynman diagrams, given below in Equation~\ref{eq:nucleon_nucleon_diagrams_sum}.
We replace spiders with sums, obtaining the following for the diagram on the left:
\begin{equation}\label{eq:feynman_amplitude_sum}
    (-ig)^2 \frac{1}{\omega_{uv}^8} \sum_{x_1, x_2}  \  \tikzfigscale{0.55}{figures/feynman_amplitude_9}
\end{equation}
We then use the following inner product:
\begin{equation}
    \tikzfigscale{0.55}{figures/feynman_amplitude_10} \ = \ e^{i 2\pi p \cdot x}
\end{equation}
to obtain a concrete expression for the amplitude associated to the diagram on the left:
\begin{equation}
\scalebox{0.9}{$
\begin{aligned}[t]
    (-ig)^2 &\sum_{x_1, x_2} \frac{1}{\omega_{uv}^8} e^{-i 2\pi p_1 \cdot x_1} e^{-i 2\pi p_2 \cdot x_2} e^{i 2\pi p'_1 \cdot x_1} e^{i 2\pi p'_2 \cdot x_2}\\
    &\sum_k \frac{1}{\omega_{ir}^4} \frac{i e^{-i  2\pi k \cdot (x_1 - x_2)}}{k^2 - m^2 + i\epsilon}
\end{aligned}
$}
\end{equation}
Simplifying the expression above, and repeating the process for the diagram on the right, we obtain the following amplitude term:
\begin{equation}
    i \mathcal{A} = (-ig)^2 \left[ \frac{i}{(p_1 - p'_1)^2 - m^2 + i\epsilon} + \frac{i}{(p_1 - p'_2)^2 - m^2 + i\epsilon} \right]
\end{equation}
This is the same amplitude computed by Feynman rules for the following traditional diagrams (see Section~\ref{appendix:feynman_diagrams} of the Appendix):
\begin{equation}
\label{eq:nucleon_nucleon_diagrams_sum}
    \mkern-24mu 
    \vcenter{\hbox{
        \scalebox{0.7}{$
        \feynmandiagram [vertical=b to a] {
            i1 
            -- [anti fermion, rmomentum'={[arrow shorten=0.3] $p'_2$}] a 
            -- [anti fermion, rmomentum'={[arrow shorten=0.3] $p_2$}] f1,
            
            a -- [scalar] b,
            
            i2 
            -- [anti fermion, rmomentum={[arrow shorten=0.3] $p'_1$}] b 
            -- [anti fermion, rmomentum={[arrow shorten=0.3] $p_1$}] f2,

            f1 -- [draw=none] y -- [draw=none] f2,
            i1 -- [draw=none] x -- [draw=none] i2,
        };
        $}
        }}
        \mkern-24mu
        +
        \mkern-24mu 
        \vcenter{\hbox{
        \scalebox{0.7}{$
        \feynmandiagram [vertical=b to a] {
            i1 
            -- [anti fermion, rmomentum'={[arrow shorten=0.3] $p'_1$}] a 
            -- [anti fermion, rmomentum'={[arrow shorten=0.3] $p_2$}] f1,
            
            a -- [scalar] b,
            
            i2 
            -- [anti fermion, rmomentum={[arrow shorten=0.3] $p'_2$}] b 
            -- [anti fermion, rmomentum={[arrow shorten=0.3] $p_1$}] f2,

            f1 -- [draw=none] y -- [draw=none] f2,
            i1 -- [draw=none] x -- [draw=none] i2,
        };
        $}
        }}
        \mkern-24mu
\end{equation}

\subsection{Composing Feynman diagrams}
\label{section:composing_feynman_diagrams}

As a calculation tool, traditional Feynman diagrams are inherently non-compositional. 
The diagrams themselves can be composed as graphs, by gluing, but the rules that turn them into amplitudes don't respect---at least, not directly---this compositional structure.
This is because Feynman rules pertain to the computation of amplitudes for specific initial and final states: in order to compose them sequentially, for example, we must integrate the individual amplitudes for all possible intermediate particle states, for all possible ways of connecting intermediate legs.

Categorical Feynman diagrams, on the other hand, are prima-facie compositional: they are morphisms in a symmetric monoidal category, $\starComplexs$-linear endomorphism of a certain hyperfinite-dimensional Hilbert space.
Instead of representing amplitudes, each categorical diagram directly embodies the full process that turns initial and final states into amplitudes: connecting the output wire of one diagram to the input wire of the other amounts exactly to integration over all possible intermediate states.
In its simplicity, however, sequential composition of categorical Feynman diagrams turns out to be surprising: because of sliding and commutation rules, composition automatically generates all possible ways of connecting intermediate legs in the 2-dimensional diagrams.

To exemplify this phenomenon, we consider the composition of two 2nd order nucleon-nucleon scattering diagrams:
\begin{equation}
    \tikzfigscale{0.55}{figures/composition_nucleon_scattering}
\end{equation}
We can use the sliding rules to linearise the resulting diagram:
\begin{align}
    &\hspace{2mm}\tikzfigscale{0.52}{figures/composition_nucleon_scattering2}\nonumber\\[1em]
    &\tikzfigscale{0.52}{figures/composition_nucleon_scattering3}
\end{align}
We repeatedly use Equation~\ref{eq:nucleon_commutation_relation} to commute annihilation operators to the left, introducing Feynman propagators in the process.
When we're done, we have obtained all possible ways of joining the external lines of the 2-dimensional Feynman diagrams that we composed, as shown by Figure~\ref{fig:composition_calculation} at the start of this work.

\subsection{Conclusions and Future Work}

In Section~\ref{section:main_contribution}, we have shown how Feynman diagrams and amplitude calculations can be accommodated within the ever broadening universe of categorical diagrammatics.
This is a significant development, in that it paves the way for the application of such techniques to a variety of important areas of computer science, from the quantum simulation of particle physics to design and compilation in the context of photonic quantum computing.

In demonstrating this applicability, we have also provided the blueprint for a variety of works to follow: extensions of the categorical and diagrammatic primitives to Fermions and vector bosons, development of specialised calculi for photonic and continuous-variable quantum computation, translation of diagrams into computable quantum circuits (e.g. for simulation purposes).

\section*{Acknowledgements} 
We would like to thank Lia Yeh and Nicola Pinzani for many intriguing discussions.

\bibliographystyle{ACM-Reference-Format}
\bibliography{references}

\appendix

\section{Feynman diagrams from Wick's Theorem}
\label{appendix:feynman_diagrams}

\subsection{Wick's Theorem}

The calculation of scattering amplitudes requires the evaluation of time ordered products such as $T\{H_I(x_1) \dots H_I(x_n)\}$.
Wick's theorem \cite{Wick1950CollisionMatrix} provides an algorithm to convert such time ordered products to sums of \emph{normal ordered} products, which are easier to evaluate.

Given a string of operators $\phi_1(x_1) \dots \phi_n(x_n)$, their normal ordered product $:\phi_1(x_1) \dots \phi_n(x_n):$ is an arrangement where all annihilation operators are placed to the right of all creation operators: first we remove all incoming particles from the state---collapsing to the zero vector, if some required mode is present in insufficient number---and then we add back all outgoing ones.

Wick's theorem is based on a simple but clever observation: that the only obstacle to time-ordering $T \{\phi(x) \phi(y)\}$ of two field operators is the commutation of positive frequency components of a field across negative frequency components \emph{of the same field}.
Expanding $\phi (x) = \phi^{+} (x) + \phi^{-} (x) $ shows that the only commutation necessary when $x^0 > y^0$ is the one below:
\begin{equation}
    \phi^{+} (x) \phi^{-} (y) = \phi^{-} (y) \phi^{+} (x) + D(x-y)
\end{equation}
From this, we get an expression for the time-ordered product in the case where $x^0 > y^0$:
\begin{equation}
    T \{\phi(x) \phi(y)\} = \ : \phi(x) \phi(y) : + \ D(x-y)
\end{equation}
Analogously, for $x^0 < y^0$ we get:
\begin{equation}
    T \{\phi(x) \phi(y)\} = \ : \phi(x) \phi(y) : + \ D(y-x)
\end{equation}
Putting the two cases together, a Feynman propagator appears:
\begin{equation}
    T \{\phi(x) \phi(y)\} = \ : \phi(x) \phi(y) : + \ \Delta_F (x-y)
\end{equation}
Analogously, for the case of complex scalar fields:
\begin{equation}
    T\{\psi(x) \psi^{\dagger}(y)\} = \ : \psi(x) \psi^{\dagger}(y) : + \ \Delta_F (x-y)
\end{equation}
The expression above represents an atomic step in the algorithm from Wick's Theorem: normal ordering of a pair of fields by introduction of a Feynman propagator.
Because such a pair can appear anywhere in a string of field operators---typically involving multiple particle types---Wick introduces \emph{contractions}, a notation to keep track of the pairs that have been normally ordered at any given step of the computation.
The contraction of $\phi(x_m)$ and $\phi(x_n)$ in a string of operators $\dots \phi(x_m) \dots \phi(x_n) \dots$ is denoted as follows:
\begin{equation}
    \wick{\dots \c{\phi(x_m)} \dots \c{\phi(x_n)} \dots}
\end{equation}
It means that the fields no longer appear in the string, and have instead been replaced by a Feynman propagator.
For example, for a product of two real scalar field operators:
\begin{equation}
    \wick{\c{\phi(x)} \c{\phi(y)}} = \Delta_F (x-y)
\end{equation}
Analogously, for a product of complex scalar field operators:
\begin{equation}
\begin{aligned}
    \wick{\c2{\psi(x)} \c2{\psi^{\dagger}(y)}} &= \Delta_F (x-y)\\
    \wick{\c2{\psi(x)} \c2{\psi(y)}} &= 0 = \wick{\c2{\psi^{\dagger}(x)} \c2{\psi^{\dagger}(y)}}
\end{aligned}
\end{equation}
Wick's Theorem~\cite{Wick1950CollisionMatrix}, finally, states that:
\begin{equation}\label{eq:wicks_theorem}
\scalebox{0.9}{$
    T\{\phi(x_1) \dots \phi(x_n)\} = \ : \phi(x_1) \dots \phi(x_n) : + :\text{all contractions}:
$}
\end{equation}
The above is known as \emph{Wick's expansion}.
As an example, we apply Wick's theorem to four real scalar fields $\phi_1 \phi_2 \phi_3 \phi_4$:
\begin{equation}
\begin{aligned}
    T\{\phi_1 \phi_2 \phi_3 \phi_4\} =& \ :\phi_1 \phi_2 \phi_3 \phi_4:\ \\
    &+ :\wick{\c {\phi_1} \c {\phi_2} {\phi_3} {\phi_4}}: + :\wick{\c {\phi_1} {\phi_2} \c {\phi_3} {\phi_4}}: + :\wick{\c {\phi_1} {\phi_2} {\phi_3} \c {\phi_4}}: \\
    &+ :\wick{{\phi_1} \c {\phi_2} \c {\phi_3} {\phi_4}}: + :\wick{{\phi_1} \c {\phi_2} {\phi_3} \c {\phi_4}}: + :\wick{{\phi_1} {\phi_2} \c {\phi_3} \c {\phi_4}}:\\
    &+ :\wick{\c1 {\phi_1} \c1 {\phi_2} \c1 {\phi_3} \c1 {\phi_4}}: + :\wick{\c1 {\phi_1} \c2 {\phi_2} \c1 {\phi_3} \c2 {\phi_4}}: + :\wick{\c2 {\phi_1} \c1 {\phi_2} \c1 {\phi_3} \c2 {\phi_4}}:
\end{aligned}
\end{equation}

\subsection{Feynman diagrams}

Wick's theorem simplifies the calculation of scattering amplitudes, but the process of evaluating integrals is repetitive and tedious.
Feynman~\cite{Feynman1949QED} introduced a diagrammatic shorthand, Feynman diagrams, with simple rules for the computation of amplitudes.

Feynman diagrams arise directly from the terms in Wick's expansion.
For example, consider the interaction Hamiltonian for scalar Yukawa theory:
\begin{equation}
    H_{\text{int}} = g \sum_{\vec{x}} \frac{1}{\omega_{uv}^3} \psi^{\dagger} (\vec{x}) \psi (\vec{x}) \phi (\vec{x})
\end{equation}
where $\psi$ is a \emph{nucleon}, $\bar{\psi}$ the corresponding \emph{anti-nucleon} and $\phi$ is the \emph{meson} mediating the ``strong force'' between them. 
The 3-operator term from the interaction gives rise to a trivalent \emph{interaction vertex}, with lines corresponding to the particle types involved:
\begin{equation}
    \vcenter{\hbox{\scalebox{0.7}{$
    \feynmandiagram[horizontal= i to int]{
            i -- [scalar, edge label = $\phi$] int,
            o1 -- [fermion, edge label = $\psi^{\dagger}$] int 
            -- [fermion, edge label' = $\psi$] o2
        };
    $}}}
\end{equation}
Each normally ordered term in Wick's expansion has a corresponding Feynman diagram, defined by taking:
\begin{itemize}
    \item input legs corresponding to annihilation operators
    \item output legs corresponding to creation operators
    \item intermediate legs (connecting vertices) corresponding to contractions
\end{itemize}
For example, let's consider nucleon scattering $NN \rightarrow NN$ in scalar Yukawa theory.
The only 2nd order term in Wick's expansion which doesn't vanish is:
\begin{equation}\label{eq:example_nucleon_nucleon}
    : \psi^{\dagger} (x_1) \psi (x_1) \psi^{\dagger} (x_2) \psi (x_2):
    \wick{\c {\phi (x_1)} \c {\phi (x_2)}}
\end{equation}
The term corresponds to the following Feynman diagram:
\begin{equation}
    \vcenter{\hbox{\scalebox{0.7}{$
    \feynmandiagram [vertical=b to a] {
        i1 
        -- [anti fermion, rmomentum'={[arrow shorten=0.3] $p'_2$}] a 
        -- [anti fermion, rmomentum'={[arrow shorten=0.3] $p_2$}] f1,
        
        a -- [scalar] b,
        
        i2 
        -- [anti fermion, rmomentum={[arrow shorten=0.3] $p'_1$}] b 
        -- [anti fermion, rmomentum={[arrow shorten=0.3] $p_1$}] f2,

        f1 -- [draw=none] y -- [draw=none] f2,
        i1 -- [draw=none] x -- [draw=none] i2,
    };
$}}}
\end{equation}
Because there are two ways of annihilating the two indistinguishable mesons in the initial state, the full interaction is actually represented by a sum of two Feynman diagrams:
\begin{equation}
    \mkern-24mu
    \vcenter{\hbox{
    \scalebox{0.7}{$
    \feynmandiagram [vertical=b to a] {
        i1 
        -- [anti fermion, rmomentum'={[arrow shorten=0.3] $p'_2$}] a 
        -- [anti fermion, rmomentum'={[arrow shorten=0.3] $p_2$}] f1,
        
        a -- [scalar] b,
        
        i2 
        -- [anti fermion, rmomentum={[arrow shorten=0.3] $p'_1$}] b 
        -- [anti fermion, rmomentum={[arrow shorten=0.3] $p_1$}] f2,

        f1 -- [draw=none] y -- [draw=none] f2,
        i1 -- [draw=none] x -- [draw=none] i2,
    };
    $}
    }}
    \mkern-24mu
    +
    \mkern-24mu
    \vcenter{\hbox{
    \scalebox{0.7}{$
    \feynmandiagram [vertical=b to a] {
        i1 
        -- [anti fermion, rmomentum'={[arrow shorten=0.3] $p'_1$}] a 
        -- [anti fermion, rmomentum'={[arrow shorten=0.3] $p_2$}] f1,
        
        a -- [scalar] b,
        
        i2 
        -- [anti fermion, rmomentum={[arrow shorten=0.3] $p'_2$}] b 
        -- [anti fermion, rmomentum={[arrow shorten=0.3] $p_1$}] f2,

        f1 -- [draw=none] y -- [draw=none] f2,
        i1 -- [draw=none] x -- [draw=none] i2,
    };
    $}
    }}
    \mkern-24mu
\end{equation}
In the diagram above, a nucleon with momentum $p_1$ interacts with a nucleon of momentum $p_2$ by exchanging a virtual meson, with two possible orderings for the outgoing particles.

As a final example, consider the following 4th order term in Wick's expansion, for the same $NN \rightarrow NN$ in scalar Yukawa theory:
\begin{equation}
\scalebox{0.8}{$
    : \psi^{\dagger} (x_1) \psi (x_1) \psi^{\dagger} (x_2) \psi (x_2):
    \wick{\c1 {\phi (x_1)} \c2 {\phi (x_2)} \c1 {\phi (x_3)} \c2 {\phi (x_4)}}
    \wick{\c3 {\psi^{\dagger} (x_3)} \c2 {\psi (x_3)} \c2 {\psi^{\dagger} (x_4)} \c3 {\psi (x_4)}}
$}
\end{equation}
One of the Feynman diagrams associated with this term is depicted below:
\begin{equation}
    \vcenter{\hbox{\scalebox{0.7}{$
    \feynmandiagram [vertical=d to a] {
        i1 
        -- [fermion, momentum={[arrow shorten=0.3] $p_2$}] a 
        -- [fermion, momentum={[arrow shorten=0.3] $p'_2$}] f1,
        
        a -- [scalar] b,
        b -- [plain, half left, looseness=1.75] c -- [anti fermion, half left, looseness=1.75] b,
        c -- [scalar] d,

        i2 
        -- [fermion, momentum'={[arrow shorten=0.3] $p_1$}] d
        -- [fermion, momentum'={[arrow shorten=0.3] $p'_1$}] f2,
        
        i1 -- [draw=none] x1 -- [draw=none] x2 -- [draw=none] x3 -- [draw=none] i2,
        f1 -- [draw=none] y1 -- [draw=none] y2 -- [draw=none] y3 -- [draw=none] f2,
    };
    $}}}   
\end{equation}
This diagram contains a loop, and is therefore known as a \emph{loop} diagram; diagrams not containing loops are known as \emph{tree} diagrams.
Loop diagrams display a pesky tendency to lead to infinities in computation, requiring renormalisation to attain sensible results.

\subsection{Feynman rules}

The algorithm for translating Feynman diagrams into scattering amplitudes is known as \emph{Feynman rules}, which we now describe for our running example of scalar Yukawa theory.
In general, a scattering amplitude takes the following form:
\begin{equation}
    i \mathcal{A} \delta (p_F - p_I)
\end{equation}
where $p_I$ and $p_F$ are the sum of initial and final particle 4-momenta respectively.
The term $\delta (p_F - p_I)$ enforces conservation of 4-momentum, but does not otherwise contribute to the scattering: the interesting term is the amplitude term $i \mathcal{A}$ itself\footnote{The factor of $i$ is added as convention, to match amplitudes in non-relativistic quantum mechanics.}.
The following Feynman rules describe how to compute it:
\begin{enumerate}[label=(\alph*)]
    \item Add a momentum variable $k$ to each internal line.
    \item Impose momentum conservation equations at each vertex.
    \item For each vertex, write down a factor of $(-ig)$.
    \item For each internal line with momentum $k$ include the following factor, where $m$ and $M$ are the masses of meson and nucleon, respectively:
    \begin{equation}
    \begin{cases}
        \frac{i}{k^2 - m^2 + i\epsilon} & \text{ for a meson}\\
        \frac{i}{k^2 - M^2 + i\epsilon} & \text{ for a nucleon}
    \end{cases}
    \end{equation}
    \item For each internal loop with the undetermined momentum $l$, integrate $l$ over 4-momentum space:
    \footnote{Particles on internal edges are virtual, so their energy is not constrained by the relativistic dispersion relation.}
    \begin{equation}
        \sum_l \frac{1}{\omega_{ir}^4} 
    \end{equation}
\end{enumerate}
We now consider a few practical examples.
To start with, we look at a 2nd order diagram for nucleon-nucleon scattering:
\begin{equation}
    \vcenter{\hbox{\scalebox{0.7}{$
    \feynmandiagram [vertical=b to a] {
        i1 
        -- [anti fermion, rmomentum'={[arrow shorten=0.3] $p'_2$}] a 
        -- [anti fermion, rmomentum'={[arrow shorten=0.3] $p_2$}] f1,
        
        a -- [scalar] b,
        
        i2 
        -- [anti fermion, rmomentum={[arrow shorten=0.3] $p'_1$}] b 
        -- [anti fermion, rmomentum={[arrow shorten=0.3] $p_1$}] f2,

        f1 -- [draw=none] y -- [draw=none] f2,
        i1 -- [draw=none] x -- [draw=none] i2,
    };
$}}}
\end{equation}
Applying the Feynman rules, we get the corresponding amplitude term:
\begin{equation}
    i\mathcal{A} = (-ig)^2 \frac{i}{(p_1 - p'_1)^2 - m^2 + i\epsilon}
\end{equation}
For the 4th order loop diagram presented earlier, the rules also involve integration over undetermined momentum:
\begin{equation}
\begin{aligned}
    i\mathcal{A} =& (-ig)^4 \sum_k \frac{1}{\omega_{ir}^4} \ \frac{i}{(p_1 - p'_1)^2 - m^2 + i\epsilon} \times \frac{i}{k^2 - M^2 + i\epsilon}\\
    &\times \frac{i}{(k+p_1-p'_1)^2 - M^2 + i\epsilon} \times \frac{i}{(p'_2 - p_2)^2 - m^2 + i\epsilon} 
\end{aligned}
\end{equation}
Next, we consider a 2nd order diagram for nucleon--anti-nucleon scattering:
\begin{equation}
\vcenter{\hbox{\scalebox{0.7}{$
    \feynmandiagram [horizontal=a to b] {
        i1 -- [anti fermion, momentum'={[arrow shorten=0.3] $p_2$}] a,
        i2 -- [fermion, momentum={[arrow shorten=0.3] $p_1$}] a,
        
        a -- [scalar] b,
        
        b -- [fermion, momentum={[arrow shorten=0.3] $p'_1$}] f1,
        b -- [anti fermion, momentum'={[arrow shorten=0.3] $p'_2$}] f2,
    };
$}}}
\end{equation}
Applying the Feynman rules, we get the corresponding amplitude term:
\begin{equation}
    i \mathcal{A} = (-ig)^2 \frac{i}{(p_1+p_2)^2 - m^2 + i\epsilon}
\end{equation}
Finally, we consider the following 4th order term for meson-meson scattering:
\begin{equation}
\vcenter{\hbox{\scalebox{0.7}{$
    \feynmandiagram [horizontal=i1 to f1] {
        i1 -- [scalar, momentum={[arrow shorten=0.3] $p_2$}] b, 
        i2 -- [scalar, momentum'={[arrow shorten=0.3] $p_1$}] a, 
        d -- [scalar, momentum={[arrow shorten=0.3] $p'_2$}] f1,
        c -- [scalar, momentum'={[arrow shorten=0.3] $p'_1$}] f2,

        a -- [fermion] c -- [fermion] d -- [fermion] b -- [fermion] a,
    };
$}}}
\end{equation}
Applying the Feynman rules, we get the corresponding amplitude term:
\begin{equation}
\begin{aligned}
    i\mathcal{A} =& (-ig)^4 \sum_k \frac{1}{\omega_{ir}^4} \ \frac{i}{k^2 - M^2 + i\epsilon} \times \frac{i}{(k+p'_1)^2 - M^2 + i\epsilon}\\
    &\times \frac{i}{(k+p'_1-p_1)^2 - M^2 + i\epsilon} \times \frac{i}{(k-p'_2)^2 - M^2 + i\epsilon} 
\end{aligned}
\end{equation}

\section{Explanation of the sliding rules}
\label{appendix:sliding_rules}

Consider a particle state $\ket{\vec{n}}$, a $k$-branch split map, and an annihilation operator $a(\vec{p})$:
\begin{itemize}
    \item If we apply the annihilation operator \textit{before} the split map, we go from $\ket{\vec{n}}$ to $\ket{\vec{n}-\delta_{\vec{p}}}$, or to the zero vector if $n_{\vec{p}} = 0$.
    The split map then distributes $\ket{\vec{n}-\delta_{\vec{p}}}$ into a superposition of all possible $k$-partition of its particle content, according to $\Theta_{\vec{n}-\delta_{\vec{p}}}^k$.
    \item If instead we apply the annihilation operator \textit{after} the split map, to the $j$-th branch, it will send $\ket{\vec{i_j}}$ to $\ket{\vec{i_j}-\delta_{\vec{p}}}$, or to the zero vector if $\left(i_j\right)_{\vec{p}} = 0$.
\end{itemize}
To see that the two scenarios above result in the same exact superposition of $k$-partitioned particle contents, it suffices to note that there is a bijective correspondence between the $(\vec{i_1}, ..., \vec{i_k}) \in \Theta_{\vec{n}-\delta_{\vec{p}}}^k$ and the $(\vec{i'_1}, ..., \vec{i'_k}) \in \Theta_{\vec{n}}^k$ such that $(i'_j)_{\vec{p}} \geq 1$, given by:
\begin{equation}
    (i_l)_{\vec{q}}
    =
    \begin{cases}
    (i'_l)_{\vec{q}} - 1 & \text{ if } l = j \text{ and } \vec{p} = \vec{q}\\
    (i'_l)_{\vec{q}} \text{ otherwise}
    \end{cases}
\end{equation}
This observation translates into the following diagrammatic sliding rule for annihilation operators over split maps, valid for each branch:
\begin{equation}
    \tikzfigscale{0.6}{figures/sliding_rule_split}
\end{equation}
Taking the adjoint of this rule yields a dual sliding rule for creation operators over merge maps, again valid for each branch:
\begin{equation}
    \tikzfigscale{0.6}{figures/sliding_rule_merge}
\end{equation}

\section{More categorical Feynman diagrams}
\label{appendix:more_categorical_feynman_diagrams}

\subsection{Nucleon--anti-nucleon scattering}

Consider the following 2nd order Feynman diagram for nucleon--anti-nucleon scattering: 
\begin{equation}
    \vcenter{\hbox{\scalebox{0.7}{$
    \feynmandiagram [horizontal=a to b] {
        i1 -- [anti fermion] a,
        i2 -- [fermion] a,
        
        a -- [scalar] b,
        
        b -- [fermion] f1,
        b -- [anti fermion] f2,
    };
    $}}}
\end{equation}
Below is the corresponding normally ordered term from Wick's expansion:
\begin{equation}
    \frac{(-ig)^2}{2} \sum_{x_1, x_2} \frac{1}{\omega_{uv}^8} 
    n_+^{\dagger}(x_2) n_-^{\dagger}(x_2) n_+(x_1) n_-(x_1)
    \wick{\c {\phi(x_1)} \c{\phi(x_2)}}
\end{equation}
Below is the corresponding categorical Feynman diagram:
\begin{equation}
    \tikzfigscale{0.55}{figures/nucleon_anti_nucleon_scattering_feynman_diagram}
\end{equation}

\subsection{Meson decay}

Consider the following 1st order Feynman diagram for meson decay:
\begin{equation}
    \vcenter{\hbox{\scalebox{0.7}{$
    \feynmandiagram [horizontal=a to b] {
        a -- [scalar] b,
        
        b -- [fermion] f1,
        b -- [anti fermion] f2,
    };
    $}}}
\end{equation}
Below is the corresponding normally ordered term from Wick's expansion:
\begin{equation}
    (-ig) \sum_{x} \frac{1}{\omega_{uv}^4} n_+^{\dagger}(x) n_-^{\dagger}(x) m(x)
\end{equation}
Below is the corresponding categorical Feynman diagram:
\begin{equation}
    \tikzfigscale{0.55}{figures/meson_decay_feynman_diagram}
\end{equation}

\subsection{Meson-meson scattering}

Consider the following 4th order Feynman diagram for meson-meson scattering: 
\begin{equation}
    \vcenter{\hbox{\scalebox{0.7}{$
    \feynmandiagram [horizontal=i1 to f1] {
        i1 -- [scalar] b, 
        i2 -- [scalar] a, 
        d -- [scalar] f1,
        c -- [scalar] f2,

        a -- [fermion] c -- [fermion] d -- [fermion] b -- [fermion] a,
    };
    $}}}
\end{equation}
Below is the corresponding normally ordered term from Wick's expansion:
\begin{equation}
\scalebox{0.75}{$ \displaystyle
    \frac{(-ig)^4}{4!} \frac{1}{(\omega_{uv}^4)^4} \sum_{x_1, x_2, x_3, x_4} 
    \left(
    \begin{aligned}
        m^{\dagger}(x_3) m^{\dagger}(x_4) m(x_1) m(x_2)
        \wick{\c2{\psi(x_1)} \c2{\psi^{\dagger}(x_2)}
        \c2{\psi(x_2)} \c2{\psi^{\dagger}(x_4)}}&\\
        \wick{\c2{\psi(x_4)} \c2{\psi^{\dagger}(x_3)}
        \c2{\psi(x_3)} \c2{\psi^{\dagger}(x_1)}}&
    \end{aligned}
    \right)
$}
\end{equation}
Below is the corresponding categorical Feynman diagram:
\begin{equation}
    \tikzfigscale{0.55}{figures/meson_scattering_feynman_diagram}
\end{equation}

\section{Translating Feynman diagrams to categorical diagrams}
\label{sec:feynman_to_string}

We list here the practical steps to translate a Feynman diagram to a categorical Feynman diagram.

\noindent\textbf{Step 1:} Make parallel wires using split and merge maps.  The number of parallel wires should be the maximum of the number of incoming and outgoing external lines.
\begin{equation}
    \scalebox{0.6}{\tikzfigscale{0.8}{figures/translation_step1}}
\end{equation}
\textbf{Step 2:} For incoming external lines, draw the corresponding annihilation operators in parallel on the left.
\begin{equation}
    \scalebox{0.6}{\tikzfigscale{0.8}{figures/translation_step2}}
\end{equation}
\textbf{Step 3:} For outgoing external lines, draw the corresponding creation operators in parallel on the right.
\begin{equation}
    \scalebox{0.6}{\tikzfigscale{0.8}{figures/translation_step3}}
\end{equation}
\textbf{Step 4:} Connect the parallel field wires of creation and annihilation operators.
\begin{equation}
    \scalebox{0.6}{\tikzfigscale{0.8}{figures/translation_step4}}
\end{equation}
\textbf{Step 5:} For each interaction vertex, place a position spider.
\begin{equation}
\scalebox{0.6}{\tikzfigscale{0.8}{figures/translation_step5}}
\end{equation}
\textbf{Step 6:} For each internal line, create a propagator and connect it to the relevant vertices (spiders).
\begin{equation}
\scalebox{0.6}{\tikzfigscale{0.8}{figures/translation_step6}}
\end{equation}
\textbf{Step 7:} For vertex connecting an external line, connect the corresponding spider to the control wire of the coherently controlled creation/annihilation operator.
\begin{equation}
\scalebox{0.6}{\tikzfigscale{0.8}{figures/translation_step7}}
\end{equation}
\textbf{Step 8:} For a Feynman diagram with $n$ interaction vertices, add a factor of $\dfrac{(-ig)^n}{n!} \dfrac{1}{(\omega_{uv}^4)^n}$.
\begin{equation}
\scalebox{0.6}{\tikzfigscale{0.8}{figures/meson_scattering_feynman_diagram}}
\end{equation}

\subsection{Computing amplitudes}
\label{appendix:computing_amplitudes}

Computation of amplitudes is simply a matter of plugging in the desired initial state and testing against the desired final co-state.
Let's go back to our nucleon-nucleon scattering example:
\begin{equation}
    \tikzfigscale{0.5}{figures/nucleon_scattering_ladder_operators_feynman_diagram}
\end{equation}
We wish to compute the amplitude for the following initial and final states:
\begin{align*}
    \ket{i} &= \ket{p_1, p_2} = n_+^{\dagger}(p_1) n_+^{\dagger}(p_2) \ket{0}\\
    \ket{f} &= \ket{p'_1, p'_2} = n_+^{\dagger}(p'_1) n_+^{\dagger}(p'_2) \ket{0}
\end{align*}
We plug these initial and final states in the above categorical diagram (omitting the factor $\frac{(-ig)^2}{2} \frac{1}{\omega_{uv}^8}$ for convenience):
\begin{equation}
\begin{aligned}
    &\quad \tikzfigscale{0.5}{figures/feynman_amplitude_1}\\[1em]
    =&\quad \tikzfigscale{0.5}{figures/feynman_amplitude_2}\\[1em]
    =&\quad \tikzfigscale{0.5}{figures/feynman_amplitude_3}
    \label{eq:appendix:feynman_amplitude_3}
\end{aligned}    
\end{equation}
To simplify this last scalar diagram, we will use the commutation relation:
\begin{equation}
    \tikzfigscale{0.5}{figures/nucleon_commutation_relation}
\end{equation}
Specifically, we will apply it to the vacuum state:
\begin{equation}
    \tikzfigscale{0.5}{figures/nucleon_commutation_relation_vacuum}
\end{equation}
Using the commutation relation, we get:
\begin{align}
    &\quad \tikzfigscale{0.5}{figures/feynman_amplitude_4} \nonumber\\[1em]
    =&\quad \tikzfigscale{0.5}{figures/feynman_amplitude_5} \nonumber\\[1em]
    =&\quad \tikzfigscale{0.5}{figures/feynman_amplitude_6}
\end{align}
We perform a symmetric simplification on the final co-state, obtaining:
\begin{equation}
    2 \left( \  \tikzfigscale{0.5}{figures/feynman_amplitude_7} \ \right)
\end{equation}
We use the fact that the vacuum state is normalized and reintroduce the $\frac{(-ig)^2}{2} \frac{1}{\omega_{uv}^8}$ factor:
\begin{equation}\label{eq:appendix:feynman_amplitude_diagram}
    (-ig)^2 \frac{1}{\omega_{uv}^8} \left( \  \vcenter{\hbox{\tikzfigscale{0.5}{figures/feynman_amplitude_8}}} \ \right)
\end{equation}
The above is functionally equivalent to the corresponding sum of traditional Feynman diagrams, given below in Equation~\ref{eq:appendix:nucleon_nucleon_diagrams_sum}.
We replace spiders with sums:
\begin{equation}\label{eq:appendix:feynman_amplitude_sum}
    (-ig)^2 \frac{1}{\omega_{uv}^8} \sum_{x_1, x_2}  \  \vcenter{\hbox{\tikzfigscale{0.6}{figures/feynman_amplitude_9}}}
\end{equation}
We then use the following inner product:
\begin{equation}
    \vcenter{\hbox{\tikzfigscale{0.8}{figures/feynman_amplitude_10}}} \ = \ e^{i 2\pi p \cdot x}
\end{equation}
to obtain a concrete expression for the amplitude:
\begin{equation}
\scalebox{0.8}{$
\begin{aligned}[t]
    (-ig)^2 &\sum_{x_1, x_2} \frac{1}{\omega_{uv}^8} e^{-i 2\pi p_1 \cdot x_1} e^{-i 2\pi p_2 \cdot x_2} e^{i 2\pi p'_1 \cdot x_1} e^{i 2\pi p'_2 \cdot x_2}\\
    &\sum_k \frac{1}{\omega_{ir}^4} \frac{i e^{-i  2\pi k \cdot (x_1 - x_2)}}{k^2 - m^2 + i\epsilon}\\
    =(-ig)^2 &\sum_k \frac{1}{\omega_{ir}^4} \sum_{x_1, x_2} \frac{1}{\omega_{uv}^8} e^{i 2\pi x_1 \cdot (p'_1 - p_1 - k)} e^{i 2\pi x_2 \cdot (p'_2 - p_2 + k)} \frac{i}{k^2 - m^2 + i\epsilon}
\end{aligned}
$}
\end{equation}
The $x_1$ and $x_2$ sums give Dirac deltas, so we obtain:
\begin{align}
    &(-ig)^2 \sum_k \frac{1}{\omega_{ir}^4} \frac{i}{k^2 - m^2 + i\epsilon} \delta^{(4)}(p'_1 - p_1 - k) \delta^{(4)}(p'_2 - p_2 + k)\\
    =\ & (-ig)^2 \frac{i}{(p_1 - p'_1)^2 - m^2 + i\epsilon} \delta^{(4)}(p'_1 + p'_2 - p_1 - p_2)
\end{align}
Analogously, on the right we obtain:
\begin{equation}
    (-ig)^2 \frac{i}{(p_1 - p'_2)^2 - m^2 + i\epsilon} \delta^{(4)}(p'_1 + p'_2 - p_1 - p_2)
\end{equation}
As a consequence, we can write the scalar \eqref{eq:feynman_amplitude_diagram} as:
\begin{equation}
\begin{aligned}[t]
    (-ig)^2 \bigg[ &\frac{i}{(p_1 - p'_1)^2 - m^2 + i\epsilon}\\
    & + \frac{i}{(p_1 - p'_2)^2 - m^2 + i\epsilon} \bigg] \delta^{(4)}(p'_1 + p'_2 - p_1 - p_2)
\end{aligned}
\end{equation}
Hence the amplitude term is:
\begin{equation}
    i \mathcal{A} = (-ig)^2 \left[ \frac{i}{(p_1 - p'_1)^2 - m^2 + i\epsilon} + \frac{i}{(p_1 - p'_2)^2 - m^2 + i\epsilon} \right]
\end{equation}
This is the same amplitude computed by Feynman rules for the following traditional diagrams (see Section~\ref{appendix:feynman_diagrams} of the Appendix):
\begin{equation}
\label{eq:appendix:nucleon_nucleon_diagrams_sum}
\mkern-24mu
    \vcenter{\hbox{
        \scalebox{0.7}{$
        \feynmandiagram [vertical=b to a] {
            i1 
            -- [anti fermion, rmomentum'={[arrow shorten=0.3] $p'_2$}] a 
            -- [anti fermion, rmomentum'={[arrow shorten=0.3] $p_2$}] f1,
            
            a -- [scalar] b,
            
            i2 
            -- [anti fermion, rmomentum={[arrow shorten=0.3] $p'_1$}] b 
            -- [anti fermion, rmomentum={[arrow shorten=0.3] $p_1$}] f2,

            f1 -- [draw=none] y -- [draw=none] f2,
            i1 -- [draw=none] x -- [draw=none] i2,
        };
        $}
        }}
        \mkern-24mu
        +
        \mkern-24mu
        \vcenter{\hbox{
        \scalebox{0.7}{$
        \feynmandiagram [vertical=b to a] {
            i1 
            -- [anti fermion, rmomentum'={[arrow shorten=0.3] $p'_1$}] a 
            -- [anti fermion, rmomentum'={[arrow shorten=0.3] $p_2$}] f1,
            
            a -- [scalar] b,
            
            i2 
            -- [anti fermion, rmomentum={[arrow shorten=0.3] $p'_2$}] b 
            -- [anti fermion, rmomentum={[arrow shorten=0.3] $p_1$}] f2,

            f1 -- [draw=none] y -- [draw=none] f2,
            i1 -- [draw=none] x -- [draw=none] i2,
        };
        $}
        }}\mkern-24mu
\end{equation}

\end{document}